\documentclass[prb,onecolumn]{revtex4}
\usepackage{graphicx}
\usepackage{amsmath}
\usepackage{amssymb}
\usepackage{stmaryrd}
\usepackage{color}

\definecolor{labelkey}{cmyk}{.4,.2,0,0}

\newcommand{\be}{\begin{equation}}
\newcommand{\ee}{\end{equation}}
\newcommand{\bea}{\begin{eqnarray}}
\newcommand{\eea}{\end{eqnarray}}
\newcommand{\nn}{\nonumber}

\begin{document}

\title{Crossover between various initial conditions in KPZ growth: flat to stationary}

\author{Pierre Le Doussal}

\affiliation{
CNRS-Laboratoire de
Physique Th{\'e}orique de l'Ecole Normale Sup{\'e}rieure, 24 rue
Lhomond,75231 Cedex 05, Paris France.}

\begin{abstract}
We conjecture the universal probability distribution at large time for the one-point height
in the 1D Kardar-Parisi-Zhang (KPZ) stochastic growth universality class, with initial conditions 
interpolating from any one of the three main classes (droplet, flat, stationary) on the left, to another
on the right, allowing for drifts and also for a step near the origin. The result is obtained from a replica Bethe ansatz calculation starting from the KPZ continuum equation, together with
a "decoupling assumption" in the large time limit. Some cases are checked to be
equivalent to previously known results from other models in the same class, 
which provides a test of the method,
others appear to be new. In particular we obtain the crossover distribution between 
flat and stationary initial conditions (crossover from Airy$_1$ to Airy$_{{\rm stat}}$) in
a simple compact form. 
\end{abstract}
\maketitle


\section{Introduction}

The one-dimensional Kardar-Parisi-Zhang (KPZ) equation \cite{KPZ} describes, in the continuum, the stochastic growth of an interface, of height $h(x,t)$ at point $x \in \mathbb{R}$, as a function of time $t$
\be \label{kpzeq}
\partial_t h(x,t) = \nu \partial_x^2 h(x,t) + \frac{\lambda_0}{2}  (\partial_x h(x,t))^2 + \sqrt{D} ~ \xi(x,t)
\ee
driven by a unit white noise $\overline{\xi(x,t) \xi(x',t')}=\delta(x-x') \delta(t-t')$. 
It has a number of experimental realizations \cite{exp4,exp5,KPZCoffee,KPZSeverine} and is at the center 
of large (and growing) universality class, which contains exactly solvable models 
in discrete settings, studied both in physics and mathematics. 
Recently there was progress in finding exact solutions for the continuum KPZ equation
itself. While the scaling exponents $h \sim t^{1/3}$, $x \sim t^{2/3}$ have 
been known for a while \cite{exponent}, the present interest is to characterize the full statistics 
of the height field $h(x,t)$. The KPZ equation can be mapped to the 
continuous directed polymer (DP) in a quenched random potential, such
that $h(x,t)= \ln Z(x,t)$ is proportional to the free energy of the DP of length $t$ with one fixed endpoint 
at $x$.

Interestingly, the KPZ interface retains some memory of the initial condition, and a few main universal statistics are found to emerge at large time, depending on the type of initial conditions. 
Remarkably, these are also related to the universality of large random matrices. This was first 
obtained from discrete models in the KPZ universality class, i.e. expected to share all its (rescaled) large time properties, such as the PNG growth model \cite{png,spohn2000,ferrari1}, the TASEP particle transport model \cite{spohnTASEP,ferrariAiry,Airy1TASEP} or discrete DP models
\cite{Johansson2000,spohn2000}. Recently it was obtained more directly, from exact solutions of
the KPZ equation: on the infinite line there are three main classes
\begin{itemize}
\item {\it The droplet (or hard wedge) initial condition} (DP with two fixed endpoints) leads to height fluctuations governed at large time by the Tracy Widom (TW) distribution $F_2$,
the CDF (cumulative distribution function) of the largest eigenvalue of the GUE random matrix ensemble \cite{TW1994Airy}. It was solved by two methods (i) as a limit from an ASEP model with weak asymetry
\cite{spohnKPZEdge} leading to a rigorous derivation \cite{corwinDP,reviewCorwin} (ii) 
using the replica Bethe ansatz (RBA) method \cite{we,dotsenko} by calculating 
the integer moments of $Z=e^{h}$ from the known exact solution of the Lieb-Liniger delta Bose 
gas \cite{ll} (also derived recently from the sine-Gordon field theory \cite{sineG}). Both methods obtained the CDF for all times $t$, as a Fredholm determinant, displaying convergence to $F_2$ as $t \to +\infty$.
\item {\it The flat initial condition} (point to line DP), solved with the RBA at all times \cite{we-flat,we-flatlong,flatnumerics,cl-14} 
and at large time \cite{dotsenkoGOE}. Rigorous calculations within ASEP have not yet led to a proof of the finite time results for the KPZ equation
(see \cite{Quastelflat} for the present status of rigorous calculations). The convergence of the one-point CDF is now to $F_1$, associated with the GOE ensemble of random matrices.
\item {\it The stationary i.e. Brownian initial condition} solved first at all times using the RBA \cite{SasamotoStationary}
exhibits convergence at large time to the Baik-Rains $F_0$ distribution. Recently it was solved rigorously as a limit of discrete directed polymer models using tools from the theory of Macdonald processes \cite{BCFV}.
 \end{itemize} 
 The RBA also allowed to solve the KPZ equation on the half-line \cite{we-halfspace} which relates to the GSE random matrix ensemble \cite{TW1996GOEGSE}. 
 
Although non-rigorous (since the integer moments $\overline{Z^n}$ of the continuum KPZ equation grow too fast with $n$, as $\sim e^{c n^3}$, to determine uniquely the distribution) the RBA has shown impressive heuristic value, often preceding rigorous results, still not available in all cases. A number of the latter
 have been obtained recently as limits (e.g. $q \to 1$) from a hierarchy of (often novel) integrable discrete models (including $q$-TASEP, $q$-bosons, semi-discrete DP, vertex models) and new mathematical tools (e.g. Macdonald processes)
 \cite{BorodinMacdo,BorodinQboson,Duality,povolo,BorodinQHahnBethe,Borodin6Vertex,CorwinVertexModels}.
 
Besides the three main classes, one expects universal {\it crossover classes} (also called transition classes) with initial conditions interpolating from one of the three classes at $x=-\infty$ to another one at $x=+\infty$, see e.g. Fig. 4 in Ref. \cite{reviewCorwin}. If the two limits are distinct classes, there are three possibilities as follows:
\begin{itemize}

\item {\it Droplet to stationary:} The KPZ equation with half-Brownian initial conditions
was solved for all times using the RBA \cite{SasamotoHalfBrownReplica}. Although the
precise form of the obtained kernel is different, it is found equivalent to 
the result obtained by taking the weak asymmetry limit \cite{QuastelAi2BM}
of the general result for the ASEP with half-Bernouilli initial conditions obtained in \cite{TW-ASEPHalfBrownian}.
At large time this leads to the universal {\it GUE to stationary} crossover distribution.
It admits an interesting generalization where the initial condition on the half-space is the partition
sum of an O'Connel-Yor directed polymer with $N$ layers, equivalently the highest eigenvalue of
the $GUE(N)$ Dyson Brownian motion ($N=1$ corresponds to
the half-Brownian) \cite{BCF}.

\item {\it Droplet to flat:}  We studied recently using RBA \cite{PLDCrossoverDropFlat} the transition from GUE to GOE in the KPZ equation, realized for the so-called "half-flat" intial condition, which is flat for $x<0$ and droplet-like for $x>0$.
From the "half-flat" formula obtained in \cite{we-flatlong}, we could produce a conjecture for the PDF in
the large time limit. We obtained a new formula for the transition Kernel and showed that it is equivalent
to the one obtained in Ref. \cite{BorodinAiry2to1} from a solution of the TASEP
with initial condition of particles on even sites for $x \leq 0$ and empty for $x>0$.
This is a mark of the expected universality at large time of this transition class. 
The corresponding Airy process was defined and characterized in 
Ref. \cite{BorodinAiry2to1}  and called ${\cal A}_{2 \to 1}$.

\item {\it Flat to stationary:}  At present there is no derivation of the
flat to stationary distribution directly for the KPZ equation. 
In the large time limit, the corresponding distribution was 
obtained in \cite{Borodin2speed} for TASEP with initial conditions $2 \mathbb{Z}_-$, i.e.
particles on even sites for $x \leq 0$, no particles for $x >0$, with the $M$ first particles 
endowed with a slower speed $\alpha<1$. With this setting there
is a point in the phase diagram of the model where the
crossover flat to stationary can be attained (the corresponding kernel 
is given by Eq. (5.4)-(5.8) there with
$M=1$ and $\kappa=0$, for $\alpha=1/2$). In terms of
process this is called ${\cal A}_{1 \to {\rm stat}}$. \\

\end{itemize}

The aim of this paper is to revisit these crossover classes starting from the KPZ equation 
\eqref{kpzeq} and using the replica Bethe ansatz method.
We will in particular obtain the lacking result for the third universal crossover class
in the KPZ equation, 
the {\it flat to stationary}. Note that for each class of initial condition there are two degrees of freedom
which can be varied, corresponding to two known invariances of the KPZ equation, namely (in units such
that $\lambda_0=D=2$, $\nu=1$ - see below) a shift in height by a constant $h(x,t) \to h(x,t) + \Delta$, and a tilt by a finite slope
$h(x,t) \to h(x+2 w t,t) + w x + w^2 t$, also known as Galilean invariance on the associated 
Burgers equation (the derivative of the KPZ equation). Hence it is quite natural to study
the crossover between initial conditions with different slopes $w_{L,R}$ 
and a height mismatch (step size) $2 \Delta$ on each side. By scaling the slopes and step size appropriately 
with time one obtains universal crossover distributions at large time (it then makes
sense as a crossover distribution even when the class is the same on each side). We also obtain for example 
the universal distribution for the wedge initial condition (flat to flat crossover) 
and for the Brownian to Brownian. In summary we expect our results to 
apply to any initial condition $h(x,t=0)$ which interpolates between
a left initial condition $h_{0,\rm left}(x)$ for $x<0$, and a right initial condition 
$h_{0,\rm right}(x)$, for $x>0$, each belonging to one
the three main classes, with possibly a mismatch in height, $2 \Delta$, and in slopes
$w_L-w_R$, merging within an interpolation region of size $x_0=t_0^{2/3}$, e.g. an initial condition of the form
\bea \label{intro} 
h(x,t=0) = (h_{0,\rm left}(x) + \Delta) \theta(-x) + t_0^{1/3} f(x/t_0^{2/3}) + ( h_{0,\rm right}(x) - \Delta) \theta(x)  
\eea
where $f(x)$ is a bounded function which decays to $0$ at infinity, and $\theta(x)$ the Heaviside unit step function.
We will consider the large time limit $t \gg t_0$, in which the precise form $f(x)$ of the interpolating region becomes irrelevant, and study the result as a scaling function
of the scaled parameters $\Delta/t^{1/3}$ and $w_{L,R} t^{1/3}$.

We obtain these results within the RBA using an assumption in the large time limit,
sometimes called a "decoupling assumption"
\cite{dotsenkoGOE,ps-2point,ps-npoint,dotsenko2pt,Spohn2ptnew,dotsenkoEndpoint,KPZFixedPoint}.
The method is similar from the one we applied to study the droplet to flat crossover in Ref. \cite{PLDCrossoverDropFlat}
(some results of that work are recovered here
in some limit) but is significantly more involved as it requires working with the combinatorics
of groups of replica, as pionneered by Dotsenko \cite{dotsenkoGOE}
(an approach that we will test and slightly extend here).
For the Brownian to Brownian crossover, our results take a different form, but agree
with known results, e.g. the one of \cite{SasamotoStationary}, 
which provides yet another a non-trivial check of the method. \\

The outline of the paper is as follows. In section \ref{sec:model} we recall the
model, the units and the mapping to the directed polymer. In section \ref{sec:init} 
we describe the initial conditions studied in this paper. In section \ref{sec:results}
we first recall standard results, then we summarize the main results of the present work. 
In section \ref{sec:gen} we give the definitions of the generating functions. Section \ref{sec:rep}
contains the main replica calculations. The quantum mechanical method 
and the Bethe ansatz are recalled, then in
section \ref{sec:combi} we display the combinatorics identity which is used,
leading to a general formula for the moments and generating function 
in \ref{sec:general}. The large time limit is studied in section \ref{sec:largetime} 
and using the decoupling assumption leads to an expression for the
generating function as a Fredholm determinant, involving
a kernel which is given in two equivalent forms. The section \ref{sec:res2}
details these two equivalent forms of the kernel in each of the various cases, with their limit
forms and comparison with known results. Finally, the section \ref{sec:conclu}
is the conclusion. The first Appendix details the combinatorial identity. The
next two appendices detail the calculations of the kernels associated to the standard generating
function, and the last one for the generalized generating function.

\section{Model and main results}
\label{sec:model} 

\subsection{KPZ equation, directed polymer and units} 

Let us consider the KPZ equation (\ref{kpzeq}), and define the scales
\bea
x^* = \frac{(2 \nu)^3}{D \lambda_0^2} \quad , \quad t^* = \frac{ 2 (2 \nu)^5}{D^2 \lambda_0^4} 
\quad , \quad   h^* = \frac{2 \nu}{\lambda_0}
\eea 
which we will use as units, i.e. we set $x \to x^* x$, $t \to t^* t$ and $h \to h^* h$
so that from now on $x,t,h$ are in dimensionless units and the KPZ equation becomes:
\bea \label{kpzeq2}
&& \partial_t h(x,t) =  \partial_x^2 h(x,t) + (\partial_x h(x,t))^2 + \sqrt{2} ~ \eta(x,t)  
\eea
where $\eta$ is also a unit white noise $\overline{\eta(x,t) \eta(x',t')}=\delta(x-x') \delta(t-t')$. 
As is well known the Cole-Hopf mapping solves the KPZ equation from an arbitrary
initial condition as follows. The solution at time $t$ can be written as:
\be
e^{h(x,t)} = Z(x,t) := \int dy Z_\eta(x,t|y,0) e^{h(y,t=0)}.
\ee
where here and below we denote $\int dy = \int_{-\infty}^{+\infty} dy$.
Here $Z_\eta(x,t|y,0)$ is the partition function of the continuum directed polymer in the random potential
$- \sqrt{2} ~ \eta(x,t)$ with fixed endpoints at $(x,t)$ and $(y,0)$:
\be \label{zdef} 
Z_\eta(x,t|y,0) = \int_{x(0)=y}^{x(t)=x}  Dx e^{-  \int_0^t d\tau [ \frac{1}{4}  (\frac{d x}{d\tau})^2  - \sqrt{2} ~ \eta(x(\tau),\tau) ]}
\ee
which is the solution of the (multiplicative) stochastic heat equation (SHE):
\bea \label{dp1} 
\partial_t Z = \nabla^2 Z + \sqrt{2} ~ \eta Z
\eea 
with Ito convention and initial condition $Z_\eta(x=0,t|y,0)= \delta(x-y)$. Equivalently, $Z(x,t)$ is the solution of
(\ref{dp1}) with initial conditions $Z(x,t=0)=e^{h(x,t=0)}$. We will adopt the notation (for the 
solution of the droplet initial condition started in $y$):
\bea \label{drop0} 
h_\eta(x,t|y,0) = \ln Z_\eta(x,t|y,0) 
\eea 
although it is somewhat improper (it requires a regularization, see below). 
We will sometimes omit the "environment" index $\eta$.
Here and below overbars denote averages over the white noise $\eta$. 

\subsection{Initial conditions}
\label{sec:init} 

We will study the KPZ equation (\ref{kpzeq2}) with the following
initial condition:
\bea \label{inith}
h(y,t=0) = h_0(y) = (w_L y + a_L B_L(-y) ) \theta(-y) + (- w_R y + a_R B_R(y)) \theta(y) 
\eea 
where $\theta(y)$ is the unit step Heaviside function, $B_L(y)$ and $B_R(y)$ are independent one-sided unit centered Brownians,
with $B_L(0)=B_R(0)=0$, $B_{L,R}(y)$ being defined for $y \geq 0$.
The correlator is $<B_R(y) B_R(y')>=\min(y,y')$ and similarly for $B_L(y)$. 
The parameters $w_L,w_R$ (usually chosen positive) measure the bias of
the Brownian, i.e. the slopes of the KPZ initial profile on each side. 

The parameters 
$a_L,a_R$ are chosen in $\{0,1\}^2$ to allow to study the four "solvable" cases (in fact three distinct ones, by 
symmetry). The wedge initial condition corresponds to $a_L=a_R=0$
and contains left (resp. right) half-flat initial condition as limits $w_R \to +\infty$ (resp. $w_L \to +\infty$). 
The Brownian to Brownian (with drifts) corresponds 
to $a_L=a_R=1$, and contains the stationary case as a limit (when $w_L=w_R=0$).
The flat to Brownian (with drifts) corresponds to $a_L=0,a_R=1$ and 
contains as limits the half-Brownian $w_L \to +\infty$ and half-flat $w_R \to +\infty$.
By symmetry it is also realized for $a_L=1, a_R=0$.

In addition, at little further expense in the calculation, we will be able to add a {\it step} $\Delta$ 
in the initial height, i.e. study the initial condition
\bea \label{step0} 
h_0^\Delta(y) = h_0(y) - \Delta ~  {\rm sgn}(y)   \label{step0} 
\eea 
where $h_0(y)$ is any of the above cases. With no loss of generality we will consider 
$\Delta \geq 0$, i.e. a downward step.

\subsection{Results} 

\label{sec:results} 

In order to obtain the most interesting large time limit, 
we need to scale the original slopes and position with time so that the
following rescaled parameters (denoted by hat)
\bea
\hat w_{L,R} = t^{1/3} w_{L,R} \quad , \quad  \hat x = \frac{x}{2 t^{2/3}} 
\eea 
are kept fixed as time becomes large. This is consistent with the standard KPZ scaling. 
Clearly this also contains the (less interesting) case where the limit is done instead with fixed $x,w_{L},w_{R}$,
which is equivalent to set $\hat w_{L,R} \to \infty$ and $\hat x \to 0$ in all formula below.

At large time the KPZ field grows linearly in time plus $\sim t^{1/3}$ fluctuations
\bea
h(x,t) \simeq v_\infty t + O(t^{1/3}) 
\eea 
and for the continuum KPZ solution
\footnote{Note that $v_\infty = -1/12$ 
at large time is a result of the Ito convention in (\ref{dp1})
which implies that $\overline{Z(x,t)}$ obeys the free diffusion equation: this defines the Cole-Hopf solution to the continuum KPZ equation for white noise. In presence of a regularized noise (i.e. with spatial 
correlations) $v_\infty$ becomes non-universal. However, as detailed in \cite{flatnumerics} if 
one considers $\ln (Z(t)/{\overline Z(t)})$ then $v_\infty = -1/12$ in our units}, $v_{\infty} = -1/12$. To get rid of this part linear in 
time we will, from now on redefine the KPZ field, and the DP partition sum, at all times, as
\bea
h(x,t) = - \frac{t}{12} + \tilde h(x,t) \quad , \quad Z(x,t) = e^{- t/12} \tilde Z(x,t) 
\eea 
and for notational simplicity, we will omit the tilde in what follow. 

\subsubsection{Recall of standard results} 

Let us first recall the standard results. The first is for $h_{drop}(x,t)$ corresponding
to the "droplet" or wedge initial condition $h_0(y)=- w|y|$ (i.e. here to $w_{L,R}=w$, $a_{L,R}=0$).
Strictly speaking, its exact solution at all times \cite{spohnKPZEdge,corwinDP,reviewCorwin,we,dotsenko}
is valid only for the "hard" wedge limit, i.e. $w \to +\infty$. However here we will be 
interested only in the large time limit, hence $w$ can be chosen arbitrary but fixed. 
At large time the one-point fluctuations of the height are governed by the 
GUE Tracy Widom (cumulative) distribution $F_2(s)$ as 
\bea \label{defxi}
&& h_{drop}(0,t) = \ln Z_{drop}(0,t) \simeq 
t^{1/3} \chi_2 + o(t^{1/3}) \quad , \quad {\rm Prob}(\chi_2 < \sigma) = F_2(\sigma) 
\eea  
where $F_2(\sigma)$ is given by a Fredholm determinant involving the Airy Kernel:
\bea \label{airyK} 
&& F_2(\sigma) = {\rm Det}[I - P_\sigma K_{Ai} P_\sigma]  \quad , \quad  K_{Ai}(v,v')= \int_{y>0} dy Ai(y+v) Ai(y+v')
= \frac{Ai(v) Ai'(v')-Ai'(v) Ai(v')}{v-v'}
\eea
and $P_\sigma(v)=\theta(v-\sigma)$ is the projector on $[\sigma,+\infty[$. 
Note that the solution (\ref{drop0}) is $h(x,t|y,0) \equiv h_{drop}(x-y,t)+ \ln(\frac{w}{2})$ 
(for large $w$), corresponding to a hard wedge
centered in $y$. Everywhere in this paper $\equiv$ means equivalent {\it in law}. The 
additive constant $\ln(\frac{w}{2})$ is necessary for regularization, but we will ignore below
all time-independent constants.

More generally, for droplet initial conditions, the multi-point correlation of the
field $h(x,t)$ is believed to converge \cite{ps-npoint,reviewCorwin,CorwinKPZensemble}
to the ones of the Airy$_2$ process ${\cal A}_2(\hat x)$ 
\cite{spohn2000,ferrariAiry} with, in our units \footnote{In several works, e.g. \cite{reviewCorwin,ps-npoint,QuastelSupremumAi2}, the dimensionless equation is chosen by setting
$\nu=\frac{1}{2}$, $\lambda_0=1$ and $D=1$. This is equivalent to only a change of the time,
i.e. it corresponds to the choice $t^*=2$ (i.e. $t'=2 t$ where $t$ denotes the time here
and $t'$ the time there).}:
\bea \label{h2} 
&& h(x,t) \simeq  
t^{1/3} ( {\cal A}_2(\hat x) - \hat x^2 ) + o(t^{1/3})
\eea 
where ${\cal A}_2(0) \equiv \chi_2$. Here $\simeq$ means in law, as a process as $x$ is varied.
The process ${\cal A}_2(\hat x)$ is stationary, i.e. statistically translationally 
invariant in $\hat x$, and well characterized: its correlations can be expressed in terms of, larger, Fredholm determinants in terms of the so-called extended Airy kernel. More generally, at large time
\bea
&& h(x,t|y,0) \simeq  
t^{1/3} ( {\cal A}_2(\hat x-\hat y) - (\hat x-\hat y)^2 ) + o(t^{1/3})
\eea 
where $h(x,t|y,0)$ is the droplet solution with arbitrary endpoints (\ref{drop0}). 
In terms of processes, this equivalence is only valid at either fixed $y$ or fixed
$x$. The process as $(x,y)$ are both varied is called the Airy sheet and is
not yet characterized. 

The second standard result is for the flat initial condition $h(x,t=0)=0$. There it was found \cite{we-flat,we-flatlong} that :
\bea \label{resflat} 
&& h_{flat}(0,t) = \ln Z_{flat}(0,t) = 
2^{-2/3} t^{1/3} \chi_1 + o(t^{1/3}) \quad , \quad 
 {\rm Prob}(\chi_1 < s) = F_1(s) 
\eea 
where $F_1(s)$ is the GOE Tracy Widom (cumulative) distribution. It 
is expressed as a Fredholm determinant 
\bea
F_1(s) = {\rm Det}[1 - P_{s/2} K^{GOE} P_{s/2}]  \quad , \quad K^{GOE}(v,v') = Ai(v+v')
\eea 
In that case, it is believed that the joint distribution of the heights $\{ h_{flat}(x,t) \}_x$ is
governed by the so-called Airy$_1$ stationary process ${\cal A}_1(u)$:
\bea
&& h(x,t) \simeq   
2^{1/3}  t^{1/3} {\cal A}_1(2^{-2/3} \hat x) + o(t^{1/3})  
\eea 
where ${\cal A}_1(0) = \frac{1}{2} \chi_1$. For definition and normalizations
of the Airy$_1$ process see e.g. Ref. \cite{Airy1TASEP,ferrariAiry,QuastelSupremumAi2}.

Note that there is a connection between these results. Indeed from
the definition one expects, in the large time limit:
\bea
&& h_{flat}(x,t) = \ln \int dy ~ e^{h(x,t|y,0)} \equiv 
\ln \int dy ~ e^{h(y,t|x,0)} \simeq 
t^{1/3} \max_{\hat y} ( {\cal A}_{2}(\hat y-\hat x) - (\hat y- \hat x)^2) 
\eea 
where we have used that the sets $\{ h(x,t|y,0) \} \equiv \{ h(y,t|x,0) \}$ are statistically
equivalent and that, since height fluctuations grow as $t^{1/3}$, the 
integral is dominated by its maximum. 
Since one can shift $\hat y$ by $\hat x$, 
the maximum of the Airy$_2$ process minus a parabola is given by the
Airy$_1$ process at one point
\bea
\max_{\hat y} ( {\cal A}_2(\hat y) - \hat y^2)  = 2^{-2/3} \chi_1 = 2^{1/3}  {\cal A}_1(0)
\eea 
as proved in \cite{QuastelSupremumAi2}.

\subsubsection{Main results of the present work} 

Let us summarize some of the results of the present work, more 
results, i.e. equivalent kernels, various limits, comparison with known results, 
and more cases, are presented in
Section \ref{sec:res2} and Appendix \ref{sec:genstep}. 

Here we obtain that, for the various initial conditions detailed in Section \ref{sec:init}, the
following CDF is given in the
limit of large time by a Fredholm determinant
\bea \label{cdf1} 
Prob( t^{-1/3} ( h(x,t) 
+ \frac{x^2}{4 t} ) < \sigma ) = {\rm Det}[ I - P_\sigma K P_\sigma ]  
\eea 
where the kernel $K$ takes the following forms in the various cases.
We need to define the function
\bea
{\cal B}_{w}(v) = e^{w^3/3} e^{-v w} - \int_0^\infty dy Ai(v+y) e^{w y}
= \int_{-\infty}^0 dy Ai(v+y) e^{w y}
\eea
where the second expression is only valid for $w>0$, while the 
first is valid for any real $w$. We find:

\begin{enumerate}

\item wedge initial condition ($a_L=a_R=0$)
\bea
K(v_i,v_j) &=&
\int_0^{+\infty} dy Ai(v_i + y) Ai(v_j + y)  (1- e^{- 2  (\hat w_L + \hat w_R) y}) \\
 &+&  
\int_{-\infty}^{+\infty} dy Ai(v_i + y) Ai(v_j - y) 
e^{ - 2 y \hat x } ( \theta(y) e^{-2 \hat w_L y} + \theta(-y) e^{2 \hat w_R y} )
\eea 
which interpolates between flat ($\hat w_{L,R} \to 0^+$) and droplet ($\hat w_{L,R} \to + \infty$)
initial conditions, and contains the half-flat (crossover ${\cal A}_{2 \to 1}$) as a special case 
for $\hat w_L=0^+$, $\hat w_R=+\infty$.

\item wedge-Brownian initial condition ($a_L=0$, $a_R=1$)
\bea
K(v_i,v_j) &=&
\int_0^{+\infty} dy Ai(v_i + y) Ai(v_j + y)  
 - Ai(v_j) \int_0^{+\infty} dy  Ai(v_i + y) e^{-  \hat x y} e^{-   (2 \hat w_L + \hat w_R) y} \\
&+&    {\cal B}_{\hat w_R - \hat x}(v_i) Ai(v_j)   + 
\int_{0}^{+\infty} dy Ai(v_i + y) Ai(v_j - y) 
e^{ - 2 y ( \hat x + \hat w_L) }  
\eea 
which for $\hat w_L=+\infty$ reproduces the half-Brownian case. 
In the limit $\hat w_L, \hat w_R \to 0^+$ we obtain the flat to stationary transition kernel
given in \eqref{main}, which is one main result of this paper. 

\item Brownian-Brownian initial condition ($a_L=1$, $a_R=1$)
\be
K(v_i,v_j) = 
\int_0^{+\infty} dy Ai(v_i + y) Ai(v_j + y)  
- \frac{Ai(v_i) Ai(v_j)}{\hat w_L + \hat w_R} 
 +  Ai(v_i)  {\cal B}_{\hat w_L + \hat x}(v_j)  +  {\cal B}_{\hat w_R - \hat x}(v_i)  Ai(v_j) 
\ee
which, as is shown in Section \ref{sec:res2} reproduces the known result for the
stationary case, although in a compact form (as a single Fredholm determinant)
to our knowledge not presented before.

\item Finally we display the result for the step initial condition, here for simplicity 
for $\hat w_{L,R}=0$ and $a_{L,R}=0$, i.e. a flat initial condition
plus a (descending) step of amplitude $\Delta>0$. For the large time
limit to be non-trivial we must scale the step
size as $t^{1/3}$, hence we define
\bea
\hat \Delta = \frac{\Delta}{t^{1/3}} 
\eea 
as the quantity kept fixed in the large time limit. 
In practice, since the KPZ equation has been derived, and is valid, only for small
height gradients $\partial_x h \ll 1$, we can think of smoothing
the step over a scale $\delta x \sim t^a$. The condition of small gradient
only requires $a \geq 1/3$, and we need $a < 2/3$ for our result to
hold (equivalent to $t \gg \tau \gg 1$ in the formulation \eqref{intro}).
With this scaling, the kernel reads
\bea
 K(v_i,v_j) &=& \int_{0}^{+\infty} dy 
Ai(v_i + y  ) 
( Ai(v_j+ y  ) - Ai(v_j+ y+ 4 \hat \Delta  ) e^{4 \hat x \hat \Delta} ) 
\\
&+&   
\int_{0}^{+\infty} dy Ai(v_i + y   ) 
Ai(v_j  - y  ) 
e^{- 2 y  \hat x } +    \int_{0}^{+\infty} dy 
Ai(v_i   -  y    )
Ai(v_j + 4 \hat \Delta +  y  ) 
e^{2 y  \hat x  }
e^{4 \hat x \hat \Delta} \nonumber 
\eea 
with however replacing the projector $P_{\sigma} \to P_{\sigma - \hat \Delta}$ 
in \eqref{cdf1}, see Eq. \eqref{formgen2n} and \eqref{sec:step2}
for more details. The generalization to arbitrary slopes $\hat w_{L,R}>0$ is given in the Appendix, equation \eqref{newMK3}.

Finally the result for a step on top of the Brownian-Brownian initial condition
is given in \eqref{SBB}. The result for a step on top of the wedge-Brownian
initial condition is given in \eqref{newMK16} and \eqref{newMK17}
(and includes the flat to stationary plus a step for $\hat w_{L,R}=0^+$). 

\end{enumerate} 

\subsection{Generating functions}
\label{sec:gen}

To obtain these results we will define and calculate some generating functions. 
For notational convenience we introduce a second set of rescaled parameters
\bea
\lambda := 2^{-2/3} t^{1/3}  \quad , \quad s = 2^{2/3} ( \sigma - \hat x^2)  \quad , \quad \tilde w_{L,R}= 
\lambda w_{L,R} = 2^{-2/3} \hat w_{L,R} \quad , \quad \tilde x = x/\lambda^2 = 2^{7/3} \hat x
\eea 
where the parameter $\lambda$ was introduced in Ref. \cite{we,dotsenko,we-flat}. As in these
works we define the standard generating function
\bea
g_\lambda(s) := \overline{\exp(- e^{- \lambda s} Z(x,t) )} = 1 + \sum_{n=1}^\infty (-1)^n \frac{e^{-n \lambda s}}{n!} 
\overline{Z(x,t)^n}
\eea 
where $Z(x,t)=e^{h(x,t)}$ and the second equality is only formal, as always in the RBA method for the continuum KPZ equation,
since the sum is a divergent series. Examination of this series, however, will allow to obtain
(or conjecture) the first average. In the large time limit it identifies with the CDF of
the rescaled height
\bea
\lim_{\lambda \to +\infty} g_\lambda(s) = Prob(\frac{1}{\lambda} h(x,t)   < s) 
\eea 

In view of the initial condition 
(\ref{inith}) it is natural to split, in each realization of $(\eta,B_L,B_R)$,
the DP partition sum into the set of paths starting at $x$ and ending either left or right of $y=0$, 
as 
\bea
&& Z(x,t) = Z^L+ Z^R \\
&& Z^L \equiv Z^L(x,t)  = \int_{-\infty}^0 dy Z_\eta(x,t|y,0) Z_0(y) \quad , \quad 
Z^R \equiv Z^R(x,t)  = \int_{0}^{+\infty} dy Z_\eta(x,t|y,0) Z_0(y) \label{LR} 
\eea 
with $Z_0(y)=Z(x,t=0)$,
and to introduce the corresponding generalized generating function:
\bea \label{genfunctLR} 
g_\lambda(s_L,s_R) = \overline{e^{-e^{-\lambda s_L} Z^L - e^{-\lambda s_R} Z^R}} 
= \sum_{n_L,n_R \geq 0}^\infty (-1)^{n_L+n_R}  \frac{e^{-n_L \lambda s_L- n_R \lambda s_R}}{n_L! n_R!} 
\overline{(Z^L)^{n_L}} \overline{(Z^R)^{n_R}}
\eea 
the standard generating function being recovered for equal arguments
\bea
g_\lambda(s)=g_\lambda(s_L=s,s_R=s)
\eea 
This generalized generating function allows to study the initial condition (\ref{step0}) in presence of
a step, which can be written as
\bea
&& Z^\Delta_0(y) = Z_0(y) e^{\Delta} \theta(-y) + Z_0(y) e^{- \Delta} \theta(y) 
\eea 
Hence the standard generating function for this problem, denoted $g_\lambda^\Delta(s)$, 
can be expressed as
\bea
g_\lambda^\Delta(s) = g_\lambda^{\Delta=0}(s_L = s - \frac{\Delta}{\lambda} , s_R = s + \frac{\Delta}{\lambda}) 
\eea

\medskip

\section{Replica calculations} 
\label{sec:rep} 

\medskip

\subsection{Averaging and quantum mechanics} 

The initial condition for the DP partition sum $Z(x,t)$ is:
\bea \label{inith}
Z(y,t=0) = Z_0(y) = e^{h_0(y)} = e^{w_L y + a_L B_L(-y)} \theta(-y) + e^{- w_R y + a_R B_R(y)} \theta(y) 
\eea 
The solution of the SHE with this initial condition can be written:
\bea \label{soluZ} 
Z(x,t) = \int dy Z_\eta(x,t|y,0) Z_0(y) 
\eea 
and we will calculate its positive integer moments with respect to the joint measure on $\eta$ and
$(B_L,B_R)$, denoted here more explicitly by overline $\overline{\cdots}^\eta$ and bracket $<..>_{B_L,B_R}$ respectively
\bea
{\cal Z}_n := \overline{<Z(x,t)^n>_{B_L,B_R}}^\eta
\eea 
Since we have chosen $\eta$ and $B_L,B_R$ to be independent it can be written as
\bea \label{zn1} 
{\cal Z}_n = \int dy_1..dy_n \overline{\prod_{\alpha=1}^n Z_\eta(x,t|y_\alpha,0)}^\eta 
< \prod_{\alpha=1}^n Z_0(y_\alpha) >_{B_L,B_R}
\eea 

Let us recall the STS symmetry. Using appendix A of Ref. \cite{we-flatlong}
one easily sees that for all $n$
\bea
\overline{<Z^n_{w_L,w_R}(x,t)>_{B_L,B_R}}^{\eta} = e^{- \frac{n x^2}{4 t} } \overline{<Z^n_{w_L+\frac{x}{2 t},
w_R- \frac{x}{2 t}}(0,t)>_{B_L,B_R}}^{\eta} 
\eea 
equivalently
\bea
\ln Z_{w_L,w_R}(x,t) + \frac{x^2}{4 t} \equiv_{in law} \ln Z_{w_L+ \frac{x}{2 t},w_R- \frac{x}{2 t}}(0,t)
\eea 
The STS symmetry thus also fixes how the generating function depends on
some combination of variables:
\bea
g_\lambda(s;w_L,w_R,x) = \tilde g_\lambda(s + \frac{\tilde x^2}{16} ; \tilde w_L + \frac{\tilde x}{8},
 \tilde w_R - \frac{\tilde x}{8}) \label{STS2} 
\eea 

As is now well known \cite{kardareplica,bb-00} the $\eta$ average in the middle of 
(\ref{zn1}) can be rewritten as 
the expectation value between initial and final states of the quantum-mechanical evolution 
operator associated to the attractive Lieb-Liniger (LL) Hamiltonian for $n$ identical particles \cite{ll}: 
\be
H_n = -\sum_{\alpha=1}^n \frac{\partial^2}{\partial {x_\alpha^2}}  - 2 \bar c \sum_{1 \leq  \alpha< \beta \leq n} \delta(x_\alpha - x_\beta) \quad , \quad \bar c=1
\label{LL}
\ee
We can thus rewrite (\ref{zn1}) in quantum mechanical notations:
\bea
{\cal Z}_n = \langle x \dots x | e^{-t H_n} | \Phi_0 \rangle 
\eea 
where $| x \dots x \rangle$ is the state will all particles at the same point $x$.
Since this state is fully symmetric in exchanges of particles, only
symmetric eigenfunctions will contribute and we can consider particles as bosons.
The wavefunction of the initial replica state is:
\bea
&& \Phi_0(Y) = \langle y_1,..y_n | \Phi_0 \rangle = 
< \prod_{\alpha=1}^n ( e^{w_L y_\alpha} e^{ a_L B_L(- y_\alpha) }  \theta(-y_\alpha) + e^{-w_R y_\alpha}  e^{ a_R B_R(y_\alpha) } \theta(y_\alpha) ) >_{B_L,B_R}  
\eea 
where here and below coordinate multiplets are denoted by capital letters, e.g. $Y \equiv y_1,..y_n$. 
This state is also clearly symmetric in the replica, hence the argument about bosons can also be
made with this state alone. 

We now introduce the decomposition into eigenstates
$\Psi_{\mu}$ and eigenenergies $E_\mu$ of $H_n$ and rewrite the moment as
a sum over eigenstates
\bea \label{sumf}
{\cal Z}_n = \sum_\mu \Psi_\mu(x,..x) \frac{e^{-t E_\mu}}{||\mu||^2}  \langle \Psi_\mu| \Phi_0 \rangle 
= \sum_\mu \Psi^*_\mu(x,..x) \frac{e^{-t E_\mu}}{||\mu||^2}  \langle \Phi_0 | \Psi_\mu \rangle
\eea 
where we have used that ${\cal Z}_n$ is real, and for convenience we will
work with the second (i.e. complex conjugate) expression. 

We can now use the explicit form of the eigenfunctions known from the Bethe ansatz \cite{ll}. They are parameterized by a set of rapidities
$\mu \equiv \{ \lambda_1,..\lambda_n\}$ which are solution of a set of coupled equations, the Bethe equations
(see below). The eigenfunctions are totally symmetric in the $x_\alpha$, and in the sector $x_1 \leq x_2 \leq \dots \leq x_n$,
take the (un-normalized) form 
\be \label{def1}
\Psi_\mu(x_1,..x_n) =  \sum_{P \in S_n} A_P \prod_{j=1}^n e^{i \sum_{\alpha=1}^n \lambda_{P_\alpha} x_\alpha} \, , \quad 
A_P=\prod_{1 \leq \alpha < \beta \leq n} \Big(1 + \frac{i  
}{\lambda_{P_\beta} - \lambda_{P_\alpha}}\Big)\,.
\ee
They can be deduced in the other sectors from their full symmetry with respect to particle exchanges. 
The sum runs over all $n!$ permutations $P$ of the rapidities $\lambda_\alpha$. The corresponding eigenenergies are
$E_\mu=\sum_{\alpha=1}^n \lambda_\alpha^2$. In the formula (\ref{sumf}) we first need: 
\be 
\Psi^*_\mu(x,..x) = n! e^{-  i x \sum_\alpha \lambda_\alpha} \,.
\label{Psixeq}
\ee 
and second, we need the overlap.
Since both states are symmetric, their overlap can be rewritten as:
\bea
&& \langle \Phi_0 | \Psi_\mu \rangle = n! \int_{ y_1<y_2< ..<y_n } \Psi_\mu(Y) \Phi_0(Y) 
 = n! \sum_{P \in S_n} A_P \int_{y_1<y_2< ..<y_n } e^{i \sum_{\alpha=1}^n \lambda_{P_\alpha} y_\alpha} \Phi_0(Y)
\eea 
using the explicit form of the Bethe eigenstates.

Introducing the numbers of replica $n_{L,R}$ on each side of $y=0$, it can be expressed
as:
\bea \label{nLnR} 
\langle \Phi_0 | \Psi_\mu \rangle = n! \sum_{P \in S_n} A_P \sum_{\substack{ n_L, n_R \geq 0 \\   n_L+n_R = n  }}
~  G^L_{n_L,w_L,a_L}[\lambda_{P_1},.., \lambda_{P_{n_L}}] G^R_{n_R,w_R,a_R}[\lambda_{P_{n-n_R+1}},.., \lambda_{P_{n}}]
\eea 
with $G^L_{0,w,a}=G^R_{0,w,a}=1$. We have defined the integrals over the left and right half-axis 
\begin{eqnarray} \label{Gla1} 
&& G_{p,w,a}^L[\lambda_1,..,\lambda_p] := \int_{y_1<y_2< ..<y_p<0} 
e^{\sum_{j=1}^p (w + i \lambda_j) y_j }  <e^{\sum_{j=1}^p a B_L(-y_j)}>_{B_L} 
\\ 
&&  G_{p,w,a}^R[\lambda_1,..,\lambda_p]  := \int_{0<y_1<y_2< ..<y_p} 
e^{\sum_{j=1}^p (- w + i \lambda_j) y_j } <e^{\sum_{j=1}^p a B_R(y_j)}>_{B_R}
 \nonumber 
\end{eqnarray}
We have taken advantage that $B_L$ and $B_R$ are independent and of the 
factorized form of each term in the wave-function in each sector. 
To evaluate these blocks
we now use the following averages over a one-sided Brownian, valid for
ordered coordinates, as indicated:
 \bea
&& <e^{\sum_{j=1}^p B(-y_j)}>_B = e^{- \frac{1}{2} \sum_{j=1}^p (2 j - 1) y_j} \quad , \quad y_1<..< y_p<0 \\
&& <e^{\sum_{j=1}^p B(y_j)}>_B = e^{\frac{1}{2} \sum_{j=1}^p (2 p - 2 j + 1) y_j} \quad , \quad 0<y_1<..< y_p
\eea
and the integration identities, valid in the domains where the integrals converge:
\begin{eqnarray} \label{Gla0} 
&&  \int_{y_1<y_2< ..<y_p<0} e^{\sum_{j=1}^p z_j y_j }  
= \prod_{j=1}^{p} \frac{1}{  z_1  + .. + z_{j} } \quad , \quad   \int_{0<y_1<y_2< ..<y_p} e^{\sum_{j=1}^p z_j y_j }  
= \prod_{j=1}^{p} \frac{- 1}{  z_p  + .. + z_{p-j+1} } 
 \nonumber 
 \eea
 
 It leads to:
\begin{eqnarray} \label{Gla1} 
&& G_{p,w,a}^L[\lambda_1,..,\lambda_p] = \int_{y_1<y_2< ..<y_p<0} 
e^{\sum_{j=1}^p (w + i \lambda_j) y_j - a \frac{1}{2} (2 j-1) y_j } 
= \prod_{j=1}^{p} \frac{1}{j w + i \lambda_1 + .. + i \lambda_{j} - a j^2/2 }\,,\\ 
&&  G_{p,w,a}^R[\lambda_1,..,\lambda_p]  = \int_{0<y_1<y_2< ..<y_p} 
e^{\sum_{j=1}^p (- w + i \lambda_j) y_j + a \frac{1}{2} (2 p+1-2 j) y_j } 
= \prod_{j=1}^{p} \frac{- 1}{-  j w + i \lambda_p + .. + i \lambda_{p+1-j} + a j^2/2 }\,,
 \nonumber 
\end{eqnarray}
Note that:
\bea
G^R_{p,w,a}[\lambda_1,..,\lambda_p]  = G^L_{p,w,a}[- \lambda_p,.. ,-\lambda_1] 
\eea 

For our two "solvable" cases, $a=0,1$ a "miracle" occurs upon performing the summation over the permutations, leading to a factorized form \cite{SasamotoHalfBrownReplica,we-flat,we-flatlong} 
\bea \label{miracle1}
&& H^R_{p,w,a=1}[\{\lambda_1,..\lambda_p\}] :=  \sum_{P \in S_p} A_P G^R_{p,w,a=1}[\lambda_{P_1},..,\lambda_{P_p}] = \frac{2^p}{ \prod_{j=1}^p (2 w - 1 - 2 i \lambda_j) }  \\
&&  H^L_{p,w,a=0}[\{\lambda_1,..\lambda_p\}] := \sum_{P \in S_p} A_P G^L_{p,w,a=0}[\lambda_{P_1},..,\lambda_{P_p}] =  \frac{1}{ \prod_{\alpha=1}^{p} (w+ i \lambda_\alpha)}  \prod_{1 \leq \alpha < \beta \leq p} \frac{2 w + i \lambda_\alpha + i \lambda_\beta - 1}{ 2 w+ i \lambda_\alpha + i \lambda_\beta}  \label{miracle2}
\eea
where we have introduced two new functions which depend only 
on the set of rapidities, not on their order. They obey now
\bea
H^R_{p,w,a}[\{\lambda_1,..\lambda_p\}] = H^L_{p,w,a}[\{- \lambda_1,.. ,-\lambda_p\}] 
\eea 

Note that these miracle identities allow to obtain simple expressions for the terms 
where either $n_L$ or $n_R$ is zero in (\ref{nLnR}) but (a priori) not for the general term
since there are then permutations which exchange rapidities in the left and
right groups of rapidities. These simpler cases were used to obtain solutions for
the half-flat and half-Brownian initial conditions \cite{SasamotoHalfBrownReplica,we-flat,we-flatlong} (formally obtained by taking one of the slopes to
infinity).

In the present case one does not seem able to proceed further 
without specifying the eigenfunctions \footnote{except for the case $a_L=a_R=1$, i.e. Brownian on
both sides where there is a further miracle identity \cite{SasamotoStationary}.}. We now recall that in the
large $L$ limit one can work with string eigenstates. These possess
specific properties which allow to obtain explicit expressions.
This is based on combinatorial properties which were 
first claimed by Dotsenko \cite{dotsenkoGOE}, 
and used by him in the case of the wedge (mostly for infinitesimal $w_{L,R}$). 
We will re-formulate, check, and slightly generalize these combinatorial
identities and apply them to other cases. 

\subsection{Strings and combinatorial identities} 
\label{sec:combi}

Let us recall the spectrum of $H_n$ in the limit of infinite system size,
i.e. the rapidities solution to the Bethe equations \cite{m-65} . 
A general eigenstate is built by partitioning the $n$ particles into a set of $1 \leq n_s \leq n$ bound states called {\it strings} 
each formed by $m_j \geq 1$ particles with $n=\sum_{j=1}^{n_s} m_j$. 
The rapidities associated to these states are written as 
\be\label{stringsol}
\lambda_{j, a}=k_j - \frac{i}2(m_j+1-2a) 
\ee 
where $k_j$ is a real momentum, the total momentum of the string being $K_j = m_j k_j$. 
Here, $a = 1,...,m_j$ labels the rapidities within the string $j=1,\dots n_s$. We will denote
$|\mu \rangle \equiv | {\bf k}, {\bf m} \rangle$ these strings states, labelled by the set of $k_j,m_j$, $j=1,..n_s$. 
Here and below the boldface represents vectors with $n_s$ components. 

Inserting these rapidities in (\ref{def1}) leads to the Bethe eigenfunctions of the infinite system, and their corresponding
eigenenergies:
\be \label{en} 
E_\mu= \frac{1}{12} n t + \tilde E( {\bf k}, {\bf m}) \quad , \quad 
\tilde E( {\bf k}, {\bf m})  := \sum_{j=1}^{n_s} m_j k_j^2-\frac{1}{12} m_j^3
\ee 
We have separated a trivial part of the energy, which can be eliminated by defining
\bea
{\cal Z}_n  = e^{- \frac{1}{12} n t}  \tilde {\cal Z}_n  \quad , \quad Z(x,t) = e^{- \frac{1}{12} t}
\tilde Z(x,t) \quad , \quad 
h(x,t) = -  \frac{1}{12} t + \tilde h(x,t)
\eea
i.e. leading to a simple shift in the KPZ field. We will implicitly study in the remainder of the paper
$\tilde {\cal Z}_n$, $\tilde Z(x,t)$ and $\tilde h(x,t)$ but will remove the tilde in these quantities
for notational simplicity (as already mentioned in the introduction). 

The formula for the norm
of the string states has been obtained as \cite{cc-07}:
\bea
&&  \frac{1}{||\mu ||^2} = \frac{1}{n!  L^{n_s} } \Phi({\bf k}, {\bf m}) \prod_{j=1}^{n_s} \frac1{m_j^{2}}  \\
&& \Phi({\bf k}, {\bf m}) = \prod_{1\leq i<j\leq n_s} 
\Phi_{k_i,m_i,k_j,m_j}   \, , \quad  \Phi_{k_i,m_i,k_j,m_j}=\frac{4(k_i-k_j)^2 +(m_i-m_j)^2}{4(k_i-k_j)^2 +(m_i+m_j)^2}
 \nonumber  \\
&& \label{norm}
\eea
so that the formula (\ref{sumf}) for the moments becomes for $L \to +\infty$ (provided all limits exist)
\bea
 {\cal Z}_n  = 
\sum_{n_s=1}^n \frac{1}{n_s!} \sum_{(m_1,..m_{n_s})_n} \prod_{j=1}^{n_s} \int \frac{dk_j}{2 \pi m_j}  
\Phi({\bf k}, {\bf m}) e^{- t \tilde E( {\bf k} , {\bf m})} e^{ - i \sum_{j=1}^{n_s} m_j k_j x} 
 \langle \Phi_0 | {\bf k}, {\bf m} \rangle
\eea 
where the second sum is over the set of partitions, denoted $(m_1,..m_{n_s})_n$, of the integer $n =\sum_{j=1}^{n_s} m_j$ into $n_s$ parts, with each $m_j \geq 1$.

It remains to calculate the overlap, formula (\ref{nLnR}). If the states are strings, the sum over permutations can be performed, using a general combinatoric identity which is detailed in the Appendix. The result is:
\bea
\langle \Phi_0 | {\bf k} , {\bf m}   \rangle = n! 
\sum_{\substack{ n_L, n_R \geq 0 \\   n_L+n_R = n  }}
\sum_{\substack{{\bf m}^L + {\bf m}^R = {\bf m} \\ \sum_{j=1}^{n_s} m_j^L=n_L \\ 
\sum_{j=1}^{n_s} m_j^R=n_R} }
\tilde H^L_{n_L,w_L,a_L}[{\bf k}, {\bf m}, {\bf m}^L]  \tilde H^R_{n_R,w_R,a_R}[{\bf k},{\bf m}, {\bf m}^R]  
 {\cal G}[{\bf k}, {\bf m}^{ L} , {\bf m}^{ R}] 
\eea 
for any string state. The sum corresponds to all possible ways to split rapidities 
in two groups associated to particles on $x<0$ (left) and $x>0$ (right). 
The combinatoric factor ${\cal G}$ is a complicated product of Gamma functions and given in the Appendix
\eqref{app:dots}. 
It coincides with the one obtained by Dotsenko \cite{dotsenkoGOE}. 
It will not play a role in the following as it will be set to unity in the large time
limit. The $\tilde H^{L,R}$ terms are obtained by 
evaluating $H^L$ and $H^R$ on the two following complementary
sets of rapidities
\bea \label{defHt} 
&& \tilde H^L_{n_L,w_L,a_L}[{\bf k}, {\bf m}, {\bf m}^L]= H^L_{n_L,w_L,a_L}[ \{ \lambda_{1,1},..\lambda_{1,m_1^L},..., \lambda_{n_s,1},.. \lambda_{n_s,m_{n_s}^L} \} ] \\
&& \tilde H^R_{n_R,w_R,a_R}[{\bf k}, {\bf m}, {\bf m}^R]= H^R_{n_R,w_R,a_R}[ \{ \bar \lambda_{1,1},..\bar \lambda_{1,m_1^R},..., \bar \lambda_{n_s,1},..
\bar \lambda_{n_s,m_{n_s}^R} \} ] \nonumber
\eea
where the first set contains the $n_L$ rapidities on the left, and the
second the $n_R$ on the right, with the following notation for the
string rapidities in the two groups
\bea \label{rapidities} 
&& \lambda_{j,r_j} = k_j -  \frac{i}{2} (m_j+1 - 2 r_j)   \quad 1 \leq r_j \leq m_j^L \\
&& \bar \lambda_{j,r_j} = k_j +  \frac{i}{2} (m_j+1 - 2 r_j)   \quad 1 \leq r_j \leq m_j^R \nonumber 
\eea 
The functions $\tilde H^{L,R}$ are thus only functions of the set of $\{ k_j,m_j^L,m_j^R\}_{j=1,..n_s}$, equivalently
$\{ k_j,m_j^L,m_j\}_{j=1,..n_s}$ or $\{ k_j,m_j,m_j^R\}_{j=1,..n_s}$, since $m_j^L+m_j^R=m_j$. 
They satisfy the symmetry relations
\bea
\tilde H^R_{n_R,w_R,a_R}[{\bf k}, {\bf m}, {\bf m}^R]= \tilde H^L_{n_R,w_R,a_R}[-{\bf k}, {\bf m}, {\bf m}^R]
= ( \tilde H^L_{n_R,w_R,a_R}[{\bf k}, {\bf m}, {\bf m}^R] )^*
\eea 

\subsection{calculation of the factors $\tilde H$}

We can now evaluate the functions $\tilde H^{L,R}$ by injecting the string rapidities (\ref{rapidities}), into the
formula (\ref{miracle1}) and (\ref{miracle2}), according to the rule (\ref{defHt}). This leads to expressions
involving Pochhammer symbols, equivalently Gamma functions. 
We have:
\bea
&& \tilde H^L_{n_L,w_L,a_L}[{\bf k} , {\bf m} , {\bf m}^L] = \prod_{j=1}^{n_s}  
S^{w_L,a_L}_{m_j^L,m_j^R,k_j} \prod_{1 \leq i < j \leq n_s} 
D^{w_L,a_L}_{m_i^L,m_i^R,k_i,m_j^L,m_j^R,k_j} \\
&& \tilde H^R_{n_R,w_R,a_R}[{\bf k} , {\bf m} , {\bf m}^R] = \prod_{j=1}^{n_s}  
S^{w_R,a_R}_{m_j^R,m_j^L,- k_j} \prod_{1 \leq i < j \leq n_s} 
D^{w_R,a_R}_{m_i^R,m_i^L,- k_i,m_j^R,m_j^L,- k_j} 
\eea 
The single string factors are:
\bea \label{single} 
&& 
S^{w,0}_{m^L,m^R,k} = \frac{(-2)^{m^L} \Gamma \left(1- 2 w - 2 i k -m^L - m^R \right)}{
\Gamma \left(1-2 w - 2 i k -m^R\right)} = 2^{m_L} \frac{\Gamma(2 w + 2 i k + m^R)}{\Gamma(2 w + 2 i k + m)}  \\
&& S^{w,1}_{m^L,m^R,k} = (-1)^{m^L} 
\frac{\Gamma(1-w - i k  - \frac{m}{2})}{\Gamma(1-w - i k  - \frac{m}{2} + m^L)} 
= \frac{\Gamma(w + i k + \frac{m}{2} - m^L)}{
\Gamma(w + i k + \frac{m}{2} )} \nn
\eea
On the first line it reproduces, for $m_R=0$, the one obtained in Ref. \cite{we-flat} (Section 5.1) for the half-flat initial condition.
The factors involving two strings are:
\be
D^{w_L,0}_{m_j^L,m_j^R,k_i,m_{j'}^L,m_{j'}^R,k_{j'}} =  \frac{\Gamma \left(1- z_{jj'} + 
\frac{1}{2} \left(-m^L_j-m^L_{j'}-m^R_j-m^R_{j'}\right)\right) \Gamma \left(1- z_{jj'} + \frac{1}{2}
   \left(m^L_j+m^L_{j'}-m^R_j-m^R_{j'}\right)\right)}{\Gamma \left(1- z_{jj'} + \frac{1}{2}
   \left(m^L_j-m^L_{j'}-m^R_j-m^R_{j'}\right)\right) \Gamma \left(1 - z_{jj'} + \frac{1}{2}
   \left(-m^L_j+m^L_{j'}-m^R_j-m^R_{j'}\right)\right)} \label{Dfact} 
\ee 
and $D^{w_L,1}_{m_j^L,m_j^R,k_i,m_{j'}^L,m_{j'}^R,k_{j'}} =1$, with 
$z_{jj'}= 2 w_L + i k_j + i k_{j'}$. In the Brownian case there is thus no 
inter-string factors. For the half-flat case, \eqref{Dfact} reproduces the one
of Ref. \cite{we-flat} (Section 5.1) when $m_j^R=m_{j'}^R=0$. 

\subsection{General formula for the moments and generating function}
\label{sec:general}

Putting all together we thus obtain the general formula for 
the moments 
\bea \label{mom1} 
&& {\cal Z}_n  = n!
\sum_{n_s=1}^n \frac{1}{n_s!}  \sum_{(m_1,..m_{n_s})_n} \prod_{j=1}^{n_s} \int \frac{dk_j}{2 \pi m_j}  \Phi({\bf k}, {\bf m}) e^{- t \tilde E( {\bf k} , {\bf m})} e^{ - i \sum_j m_j k_j x} \\
&& 
\times 
\sum_{\substack{ n_L, n_R \geq 0 \\   n_L+n_R = n  }}
\sum_{\substack{{\bf m}^L + {\bf m}^R = {\bf m} \\ \sum_{j=1}^{n_s} m_j^L=n_L \\  \sum_{j=1}^{n_s} m_j^R=n_R} }
\tilde H^L_{n_L,w_L,a_L}[{\bf k}, {\bf m}, {\bf m}^L]  \tilde H^R_{n_R,w_R,a_R}[{\bf k},{\bf m}, {\bf m}^R]  
 {\cal G}[{\bf k}, {\bf m}^L , {\bf m}^R] \nn
\eea
Note that in this sum the term with fixed $n_L,n_R$ has a simple interpretation. 
Consider \eqref{LR}, \eqref{genfunctLR}, where in each realization of $(\eta,B_L,B_R)$,
the DP partition sum is splitted into the set of paths starting at $x$ and ending either left or right of $y=0$, 
The moments then split as:
\bea
{\cal Z}_n = \overline{(Z^L + Z^R)^n} = \sum_{n_L=0}^n \sum_{n_R=0}^n \delta_{n_L+n_R,n} 
\frac{n!}{n_L! n_R!} \overline{(Z^L)^{n_L} (Z^R)^{n_R}}
\eea 
Hence, by simple identification of the terms with fixed $n_L,n_R$ in (\ref{mom1}) 
and multiplication by the factor $\frac{n_L! n_R!}{n!}$  we obtain an
expression for each joint moment $\overline{(Z^L)^{n_L} (Z^R)^{n_R}}$. 

Let us come back to the generating function (\ref{genfunctLR}). 
The sums over the variables $m,m_L,m_R$ now become free summations leading to
\bea
&& g_\lambda(s) = 1 + \sum_{n_s \geq 1} \frac{1}{n_s!} Z(n_s,s) \\
&& g_\lambda(s_L,s_R) = 1 + \sum_{n_s \geq 1} \frac{1}{n_s!} Z(n_s,s_L,s_R) 
\eea
with
\bea
&& Z(n_s,s_L,s_R) = \prod_{j=1}^{n_s} [ \sum_{m_j=1}^\infty \int \frac{dk_j}{2 \pi m_j} 
(-1)^{m_j} ] e^{- i \sum_j m_j k_j x }
 \Phi({\bf k}, {\bf m}) e^{- t \tilde E( {\bf k} , {\bf m})} \sum_{{\bf m}^L + {\bf m}^R = {\bf m} } 
 e^{- \sum_j ( \lambda m^L_j s_L + \lambda m^R_j s_R)} \\
&& 
\times 
 \prod_{j=1}^{n_s}   S^{w_L,a_L}_{m_j^L,m_j^R,k_j} S^{w_R,a_R}_{m_j^R,m_j^L,-k_j}   \times 
\prod_{1 \leq i < j \leq n_s} 
D^{w_L,a_L}_{m_i^L,m_i^R,k_i,m_j^L,m_j^R,k_j} D^{w_R,a_R}_{m_i^R,m_i^L,- k_i,m_j^R,m_j^L,- k_j}   
 {\cal G}[{\bf k}, {\bf m}^L , {\bf m}^R] \nonumber 
\eea 
Although this is an exact and explicit expression, 
apart from the case $a_L=a_R=1$, it is unclear how to handle it for arbitrary time. We thus now turn to the large time
limit. 

\subsection{large time limit}
\label{sec:largetime} 

In the large time limit we will {\it assume} that one can set the
product of factors $D$ and ${\cal G}$ to unity. This is of course a highly non-trivial and radical assumption,
however it is justified a posteriori by the results. It will be checked in all cases where
the solution is known by other means. This procedure follows what has been done in
other works, where it was also checked against other methods\cite{dotsenkoGOE,PLDCrossoverDropFlat,ps-2point,dotsenko2pt,Spohn2ptnew}.

\subsubsection{determinantal form}

Let us first obtain a closed expression once these factors are set to unity, and take
the large $\lambda$ limit in a second stage: 
\bea
&& Z(n_s,s_L,s_R) = \prod_{j=1}^{n_s} [ \sum_{m_j=1}^\infty \int \frac{dk_j}{2 \pi m_j} 
(-1)^{m_j} e^{- i m_j k_j x } ]
 \Phi({\bf k}, {\bf m}) e^{- t \tilde E( {\bf k} , {\bf m})} \\
&& 
\times 
\sum_{{\bf m}^L + {\bf m}^R = {\bf m} } 
 \prod_{j=1}^{n_s}   S^{w_L,a_L}_{m_j^L,m_j^R,k_j} S^{w_R,a_R}_{m_j^R,m_j^L,-k_j} e^{- \lambda m^L_j s_L - \lambda m^R_j s_R}
\eea 
We now use the standard determinant double-Cauchy identity:
\bea
\Phi({\bf k}, {\bf m}) = \prod_{j=1}^{n_s} (2 m_j) \det_{1 \leq i, j \leq n_s}[ \frac{1}{2 i (k_i-k_j) + m_i + m_j}] 
\eea 
and perform the rescaling $k_j \to k_j/\lambda$. Denoting $\tilde x=x/\lambda^2$ we obtain:
\bea
&& Z(n_s,s_L,s_R) = 2^{n_s} \prod_{j=1}^{n_s} [ \sum_{m_j=1}^\infty \int \frac{dk_j}{2 \pi} 
 (-1)^{m_j} e^{- i \lambda  m_j k_j \tilde x  + \frac{1}{3} \lambda^3 m_j^3 - 4 \lambda m_j k_j^2 } ] \det_{1 \leq i, j \leq n_s} [ \frac{1}{2 i (k_i-k_j) + \lambda m_i + \lambda m_j}]  \\
&& 
\times \sum_{{\bf m}^L + {\bf m}^R = {\bf m} } 
\prod_{j=1}^{n_s}  S^{w_L,a_L}_{m_j^L,m_j^R,\frac{k_j}{\lambda}} S^{w_R,a_R}_{m_j^R,m_j^L,-\frac{k_j}{\lambda}}  
e^{- \lambda m^L_j s_L - \lambda m^R_j s_R}
 \eea
We now perform the Airy trick, i.e. use the representation 
$e^{\frac{1}{3} \lambda^3 m^3} = \int dy Ai(y) e^{\lambda m y}$ to obtain
\bea
&& Z(n_s,s_L,s_R) = 2^{n_s} \prod_{j=1}^{n_s} [ \sum_{m_j=1}^\infty \int \frac{dk_j}{2 \pi} dy_j Ai(y_j) 
 (-1)^{m_j} e^{- i \lambda  m_j k_j \tilde x  - 4 \lambda m_j k_j^2 + \lambda m_j y_j  } ]  \\
&& 
\det_{1 \leq i, j \leq n_s}[ \frac{1}{2 i (k_i-k_j) + \lambda m_i + \lambda m_j}]  \times \sum_{ {\bf m}^L + {\bf m}^R = {\bf m} } 
\prod_{j=1}^{n_s}  S^{w_L,a_L}_{m_j^L,m_j^R,\frac{k_j}{\lambda}} S^{w_R,a_R}_{m_j^R,m_j^L,-\frac{k_j}{\lambda}}     
e^{- \lambda m^L_j s_L - \lambda m^R_j s_R}
 \eea
Using standard manipulations \cite{we,dotsenko} the partition sum at fixed number of string $n_s$ 
can thus be expressed itself as a determinant:
\bea
&& Z(n_s,s_L,s_R) = 
 \prod_{j=1}^{n_s} \int_{v_j>0} \;
 \det_{1 \leq i, j \leq n_s} M_{s_L,s_R}(v_i,v_j) \\
\eea
with the Kernel:
\bea \label{Mss} 
 && M_{s_L,s_R}(v_i,v_j) = \int \frac{dk}{2 \pi} dy Ai(y + 4 k^2 + i k \tilde x + v_i + v_j) e^{- 2 i k (v_i-v_j)} 
 \phi_\lambda(k,y-s_L,y-s_R) \\
&& \phi_\lambda(k,y_L, y_R) =  2 \sum_{m^L \geq 0, m^R \geq 0,m^L+m^R \geq 1} (-1)^{m_L+m_R} 
S^{w_L,a_L}_{m^L,m^R,\frac{k}{\lambda}} S^{w_R,a_R}_{m^R,m^L,-\frac{k}{\lambda}} 
 e^{\lambda m^L y_L +  \lambda m^R y_R } 
 \eea
 where the $S$ factors are given explicitly in (\ref{single}). 
The generating function is thus a Fredholm determinant:
\bea
g_\lambda(s_L,s_R) = {\rm Det}[ I + P_0 M_{s_L,s_R} P_0 ]  
 \label{ggene2} 
\eea 
where, again, this expression is valid as soon as the factors $D$ and ${\cal G}$ are set (arbitrarily) to unity.\\

We must now study the function $\phi_\lambda(k,y_L, y_R)$ in the
large time limit $\lambda \to +\infty$. 

\subsubsection{large time limit} 

We first rewrite:
\bea
 \phi_\lambda(k,y_L, y_R) =  -2 + 2 \sum_{m^L \geq 0, m^R \geq 0} (-1)^{m^L+m^R} 
S^{w_L,a_L}_{m^L,m^R,\frac{k}{\lambda}} S^{w_R,a_R}_{m^R,m^L,-\frac{k}{\lambda}} 
 e^{\lambda m^L y_L +  \lambda m^R y_R }
 \eea
and we use the Mellin-Barnes identity:
\bea
\sum_{m=0}^\infty (-1)^m f(m) = \frac{-1}{2 i} \int_C \frac{dz}{\sin \pi z} f(z) 
\eea 
where $C=\kappa + i \mathbb{R}$, $-1<\kappa<0$, valid provided $f(z)$ is meromorphic, with no pole 
for $z> \Re(\kappa)$, and sufficient decay at infinity. It allows to rewrite (for $2 w_{L,R}+\kappa>0$)
\bea
\phi_\lambda(k,y_L, y_R) =  -2 + 2  (\frac{-1}{2 i})^2 \int_C \frac{dz_L}{\sin \pi z_L} \int_C \frac{dz_R}{\sin \pi z_R} 
S^{w_L,a_L}_{z_L,z_R,\frac{k}{\lambda}} S^{w_R,a_R}_{z_R,z_L,-\frac{k}{\lambda}} 
 e^{\lambda z_L y_L +  \lambda z_R y_R }
\eea
Here the analytic continuation $f(m) \to f(z)$ has been performed using the 
second expression in (\ref{single}) as
\bea \label{single2} 
&& 
S^{w,0}_{z_L,z_R,k} = 2^{z_L} \frac{\Gamma(2 w + 2 i k + z_R)}{\Gamma(2 w + 2 i k + z_R+z_L)}  \\
&& S^{w,1}_{z_L,z_R,k} = \frac{\Gamma(w + i k + \frac{z_R-z_L}{2})}{
\Gamma(w + i k + \frac{z_R+z_L}{2} )} \nn
\eea

We now rescale $z_{L,R} \to z_{L,R}/\lambda$, and we study the large time limit $\lambda \to +\infty$. 
We first recall the definition of the rescaled drifts:
\bea
\tilde w_L = w_L \lambda \quad , \quad \tilde w_R = w_R \lambda
\eea 
and we use that for $a=0,1$:
\bea
\lim_{\lambda \to +\infty} S^{w=\tilde w/\lambda,a}_{\frac{z_L}{\lambda},\frac{z_R}{\lambda},\frac{k}{\lambda}} 
=  1 + \frac{ (1+a) z_L}{2 \tilde w + 2 i k + z_R - a z_L}
\eea 
as can be seen from (\ref{single2}). Thus we obtain the infinite $\lambda$ limit in the form of a double
contour integral:
\bea \label{phi1} 
&& \phi_{+\infty}(k,y_L, y_R) =  -2 \\
&& + 2 \int_{C'} \frac{dz_L}{2 i \pi z_L} \int_{C'} \frac{dz_R}{2 i \pi z_R} 
(1 + \frac{ (1+a_L) z_L}{2 \tilde w_L + 2 i k + z_R - a_L z_L})(1 + \frac{ (1+a_R) z_R}{2 \tilde w_R - 2 i k + z_L - a_R z_R}) 
 e^{z_L y_L +  z_R y_R } \nn
\eea
where $C'=0^- + i \mathbb{R}$. The calculation of this integral is performed in Appendix \ref{app:1}
and the result is displayed in (\ref{phitotal}). \\

For now, we focus on the simpler generating function, i.e. we set $s_L=s_R=s$.
In the infinite time limit it takes the form
\bea \label{formgen1} 
g_{+\infty}(s) = Det[ I + P_0 M_s P_0 ]  
\eea 
where $P_0(v)=\theta(v)$ is the projector on $[0,+\infty[$ and with the Kernel
\bea \label{M2} 
&& M_s(v_i,v_j) = \int \frac{dk}{2 \pi} dy Ai(y + s+ 4 k^2 + i k \tilde x + v_i + v_j) e^{- 2 i k (v_i-v_j)} 
 \phi_{+\infty}(k,y) \\
&& - \frac{1}{2} \phi_{+\infty}(k,y) = \theta(y)  [ 1 - (1+a_L+a_R-3 a_L a_R)
e^{- ( (2 \tilde w_L + 2 i k) (1+a_R)  + (2 \tilde w_R - 2 i k) (1+a_L) ) y} ] \label{phi2} \\
&& + 2 \theta(-y) (a_L e^{(2 \tilde w_L + 2 i k) y} +a_R e^{(2 \tilde w_R - 2 i k) y}) 
+ \delta(y) [  \frac{1-a_L}{2 \tilde w_L + 2 i k }  + \frac{1-a_R}{2 \tilde w_R - 2 i k} 
- \frac{a_R a_L }{\tilde w_L + \tilde w_R}    ] \nonumber 
\eea
The function $\phi_{+\infty}(k,y):=\phi_{+\infty}(k,y_L=y,y_R=y)$ being obtained 
from the more general result (\ref{phitotal}) in the Appendix. \\

We now use Airy function identities 
in order to rewrite the result in terms of an alternative kernel.
The calculation is performed in Appendix \ref{app:Airy}.
The final result is:
\bea \label{final1} 
&& g_{+\infty}(s) = {\rm Det}[ I -  P_0 K_\sigma P_0 ]  
\quad , \quad K_\sigma(v_1,v_2)=
K(v_1+\sigma ,v_2+\sigma) \\
&& \sigma = 2^{-2/3} (s+ \frac{\tilde x^2}{16}) \quad , \quad \hat w = 2^{2/3} \tilde w 
\quad , \quad \hat x  = 2^{2/3} \frac{\tilde x}{8}  \\
&&  \label{final2}  K(v_1,v_2) = 
\int_0^{+\infty} dy Ai(v_i + y) Ai(v_j + y)  
- \frac{a_R a_L }{\hat w_L + \hat w_R}  Ai(v_i) Ai(v_j) \\
&& - (1+a_L+a_R-3 a_L a_R) \int_0^{+\infty} dy  Ai(v_i + (1-a_L+a_R) y) Ai(v_j + (1+a_L-a_R) y) 
\nn \\
&& \times 
 e^{- 2 y ( \hat w_L + \hat w_R + a_R (\hat w_L + \hat x) + a_L (\hat w_R - \hat x) )}  \nn \\
&& +  a_L Ai(v_i)  \int_{-\infty}^{0} dy  Ai(v_j   + y  )
 e^{(\hat x + \hat w_L) y} +  a_R Ai(v_j ) \int_{-\infty}^{0} dy Ai(v_i + y) 
 e^{(\hat w_R - \hat x ) y} \nn \\
 && + (1-a_L) 
\int_{0}^{+\infty} dy Ai(v_i + y) Ai(v_j - y) 
e^{ - 2 y ( \hat x + \hat w_L) }  + (1-a_R) 
\int_{0}^{+\infty} dy Ai(v_i - y) Ai(v_j + y) 
e^{ - 2 y (\hat w_R - \hat x ) } \nn
\eea 
which, we note depends only on $\sigma$ and the combinations $\hat w_L + \hat x$ and $\hat w_R - \hat x$
(and their sum), as required by the STS symmetry \eqref{STS2}.

\section{Results for the various crossover kernels}
\label{sec:res2} 

We now discuss in details the results for the various initial conditions. 
We give both the result in the first form \eqref{formgen1}, with kernel $M_s$ from \eqref{M2}
\bea \label{formgen1n} 
\lim_{t \to +\infty} 
{\rm Prob}\left(t^{-1/3} h(x=2^{-4/3} t^{2/3} \tilde x,t) < 2^{-2/3} s\right) = g_{+\infty}(s;\tilde x,\tilde w_L,\tilde w_R) = {\rm Det}[ I + P_0 M_{s} P_0 ] 
\eea
naturally expressed in the variables $\tilde x, \tilde w_{L,R}=2^{-2/3} t^{1/3} w_{L,R}$, 
and the second, equivalent form of the result \eqref{final1}, with kernel $K_\sigma$ and $K$ from \eqref{final2}
\bea \label{formgen2n} 
{\rm Prob}\left(t^{-1/3} (h(x=2 t^{2/3} \hat x,t) + \frac{x^2}{4 t} ) < \sigma \right) = g_{+\infty}(s) = {\rm Det}[ I -  P_0 K_\sigma P_0 ]  = {\rm Det}[ I -  P_\sigma K P_\sigma ]
\eea 
naturally expressed in the variables $\sigma, \hat x, \hat w_{L,R}=t^{1/3} w_{L,R}$.

\subsection{the wedge initial condition}

Let us start with the wedge initial condition \eqref{inith} with $a_L=a_R=0$. 

\subsubsection{first form of the wedge kernel}

From \eqref{M2} we find
\bea \label{M2wedge} 
&& M_s(v_i,v_j) = \int \frac{dk}{2 \pi} dy Ai(y + s+ 4 k^2 + i k \tilde x + v_i + v_j) e^{- 2 i k (v_i-v_j)} 
 \phi_{+\infty}(k,y) \\
&& \phi_{+\infty}(k,y) = -2 \theta(y)  [ 1 - 
e^{- 2 (\tilde w_L + \tilde w_R) y} ]  - \delta(y) [  \frac{1}{\tilde w_L +  i k }  + \frac{1}{ \tilde w_R -  i k} 
   ] \label{phiwedge1}
\eea
Let us discuss several limits.\\

{\it Half-flat initial condition and GUE-GOE crossover:} For $\tilde w_R \to +\infty$ one recovers the Kernel for the half-flat case
obtained in Ref. \cite{PLDCrossoverDropFlat} (formula (80-81) there). 
As shown there it interpolates between the GOE (flat) for $\tilde x \to -\infty$ and the GUE (droplet) kernels
for $\tilde x \to + \infty$.\\

{\it Symmetric wedge:} In that case one chooses $\tilde w_L=\tilde w_R=w$. We obtain
\bea \label{M2wedgewsym} 
&& M_s(v_i,v_j) = \int \frac{dk}{2 \pi} dy Ai(y + s+ 4 k^2 + i k \tilde x + v_i + v_j) e^{- 2 i k (v_i-v_j)} 
 \phi_{+\infty}(k,y) \\
&& \phi_{+\infty}(k,y) = -2 \big( \theta(y) ( 1 - 
e^{- 4 \tilde w y} )  + \delta(y)  \frac{\tilde w}{\tilde w^2 + k^2} \big) \label{phiwedgesym}
\eea
This kernel also provides an interpolation from GUE (droplet) to GOE (flat) as $\tilde w$ is decreased
from $+\infty$ to $0$. The two limits are particularly immediate on that form of the kernel. For $\tilde w \to +\infty$ one has:
\bea
&& M_s(v_i,v_j) \to M_s^{GUE}(v_i,v_j) = 
 - 2 \int \frac{dk}{2 \pi} \int_0^{+\infty} dy Ai(y + s + 4 k^2 + i k \tilde x + v_i + v_j) e^{- 2 i k (v_i-v_j)} 
\eea 
This is identical to the GUE kernel in the form given in \cite{we}. In the other limit
$\tilde w \to 0^+$ we can replace
\bea
\frac{\tilde w}{\tilde w^2 + k^2} \to_{\tilde w \to 0^+} \pi \delta(k) 
\eea 
leading to:
\bea
&& M_s(v_i,v_j) \to M_s^{GOE}(v_i,v_j) = -  Ai(s + v_i + v_j) 
\eea 
which is the simplest form of the GOE kernel. 

\subsubsection{second form of the wedge kernel}

The second form of the wedge kernel reads:
\bea
&& K(v_i,v_j) = 
\int_0^{+\infty} dy Ai(v_i + y) Ai(v_j + y)  (1- e^{- 2  (\hat w_L + \hat w_R) y}) \\
 && + 
\int_{-\infty}^{+\infty} dy Ai(v_i + y) Ai(v_j - y) 
e^{ - 2 y \hat x } ( \theta(y) e^{-2 \hat w_L y} + \theta(-y) e^{2 \hat w_R y} )
\eea 

In the limit $\tilde w_R \to +\infty$ one recovers the Kernel for the half-flat case
in the second form obtained in \cite{PLDCrossoverDropFlat} (formula (89-91) there),
namely
\bea
&& K^{\rm half-flat}(v_i,v_j) = 
\int_0^{+\infty} dy Ai(v_i + y) Ai(v_j + y)  + \int_{0}^{+\infty} dy Ai(v_i + y) Ai(v_j - y) 
e^{ - 2 y (\hat w_L + \hat x) } \label{K1/2}
\eea 
which, as discussed there, is equivalent to the result of Ref. \cite{BorodinAiry2to1}
for TASEP. As shown there it interpolates between the GOE (flat) 
for $\hat x \to -\infty$ and the GUE (droplet) kernels
for $\hat x \to +\infty$, i.e. it is the (one-point) kernel associated to the ${\cal A}_{2 \to 1}$ interpolation
process. 

More generally in the double limit $(\hat w_{L},\hat w_{R}) \to (0^+,0^+)$ we obtain, using
another Airy identity:
\bea
&& K(v_i,v_j) \simeq \int_{-\infty}^{+\infty} dy Ai(v_i + y) Ai(v_j - y) e^{ - 2 y \hat x } \\
&& = 2^{-1/3} Ai( 2^{-1/3} (v_i+v_j-2 \hat x^2) e^{\hat x (v_i-v_j)} 
\eea 
Hence using that $\sigma - \hat x^2 = 2^{-2/3} s$:
\bea
&& K_\sigma(v_i,v_j) = K(v_i + \sigma ,v_j+\sigma ) \simeq 2^{-1/3} Ai( 2^{-1/3} (v_i+v_j) + s ) e^{\hat x (v_i-v_j)} 
\eea 
Under a similarity transformation this is equivalent to the GOE kernel:
\bea
K_\sigma(v_i,v_j) \equiv Ai(v_i+v_j+s) 
\eea

\subsection{the wedge-Brownian initial condition}

Let us now consider now the wedge-Brownian initial condition \eqref{inith},
with $a_L=0$ and $a_R=1$. This case contains the flat to stationary crossover
as a limit, see below.
\subsubsection{first form of the kernel}
From \eqref{M2} we find:
\bea \label{M3} 
&& M_s(v_i,v_j) = \int \frac{dk}{2 \pi} dy Ai(y + s+ 4 k^2 + i k \tilde x + v_i + v_j) e^{- 2 i k (v_i-v_j)} 
 \phi_{+\infty}(k,y) \\
&& \phi_{+\infty}(k,y) = - 2 \theta(y)  [ 1 - 2
e^{- 2 (2 \tilde w_L  + \tilde w_R  + i k) y} ]  - 4 \theta(-y) e^{(2 \tilde w_R - 2 i k) y}
- \delta(y) \frac{1}{\tilde w_L + i k }  
\eea

\subsubsection{second form of the kernel}

The second form of the kernel reads:
\bea \label{wb2} 
&& K(v_i,v_j) = 
\int_0^{+\infty} dy Ai(v_i + y) Ai(v_j + y)  -  Ai(v_j) \int_0^{+\infty} dy  Ai(v_i + y) 
e^{-  \hat x y}  e^{-   (2 \hat w_L + \hat w_R) y} \nn \\
&&  +  Ai(v_j ) \int_{-\infty}^{0} dy Ai(v_i + y) 
 e^{(\hat w_R - \hat x ) y}  + 
\int_{0}^{+\infty} dy Ai(v_i + y) Ai(v_j - y) 
e^{ - 2 y ( \hat x + \hat w_L) }  \nn
\eea 
where we have performed the change of variable $y \to y/2$ in the second term.
Note that the second integral is convergent only for $\hat w_R - \hat x > 0$.
It is however easily extended to arbitrary values (see below). \\

{\it Half-Brownian limit:}
in the limit $\hat w_L \to +\infty$ one should recover the half-Brownian initial condition. In that limit
\bea 
&& K(v_i,v_j) = 
\int_0^{+\infty} dy Ai(v_i + y) Ai(v_j + y)   +  Ai(v_j ) \int_{-\infty}^{0} dy Ai(v_i + y) 
 e^{(\hat w_R- \hat x) y}  
\eea 
Note that the second integral is convergent only for $\hat w_R - \hat x > 0$.
To obtain a more general expression, we can use the identity, valid for $u>0$:
\bea
\int_{-\infty}^{+\infty} dy Ai(v_i + y) e^{u y} = e^{\frac{u^3}{3} - u v_i}  \label{replace1}
\eea 
and replace 
\bea \label{hb1} 
&& K(v_i,v_j) = 
\int_0^{+\infty} dy Ai(v_i + y) Ai(v_j + y)   +  Ai(v_j ) 
( e^{\frac{1}{3} (\hat w_R - \hat x)^3} 
e^{- (\hat w_R-\hat x) v_i}  - 
\int_{0}^{\infty} dy Ai(v_i + y) 
 e^{(\hat w_R- \hat x) y}  )
\eea 
and expression where now the integrals are convergent for any $\hat w_R - \hat x$ and 
which coincides with the asymptotic large time formula (2.23) in
Ref. \cite{SasamotoHalfBrownReplica} (the correspondence is that $X,\gamma_t$ there are $X=\hat x - \hat w_R$, $\gamma_t=t^{1/3}$). Thus the above replacement (\ref{replace1}) is
legitimate (it can in fact be shown also from the first form of the kernel, repeating the
calculation of Appendix \ref{app:Airy})
and we will use it repeatedly in the following. \\


We can now go back to the general case of the wedge-Brownian initial condition
\eqref{wb2} and note that it can be written as the sum of the half-flat kernel (which interpolates
between GUE and GOE) and a projector
\bea \label{wb2} 
&& K(v_i,v_j) = K^{\rm half-flat}(v_i,v_j) + \Phi(v_i) Ai(v_j) 
\eea
where 
\bea
\Phi(v_i)  = e^{\frac{1}{3} (\hat w_R - \hat x)^3} 
e^{- (\hat w_R-\hat x) v_i}  - 
\int_{0}^{\infty} dy Ai(v_i + y) e^{- \hat x y} [ e^{\hat w_R y} + e^{-   (2 \hat w_L + \hat w_R) y} ]
\eea
and $K^{\rm half-flat}(v_i,v_j)$ is given in \eqref{K1/2}.\\

{\it Flat to stationary crossover:}
It is now possible to consider the limit $\hat w_L, \hat w_R \to 0^+$. One obtains
\bea \label{main}
K(v_i,v_j) &=&
\int_0^{+\infty} dy Ai(v_i + y) Ai(v_j + y) + 
\int_{0}^{+\infty} dy Ai(v_i + y) Ai(v_j - y) 
e^{ - 2 y  \hat x }  
 \\
&+&    Ai(v_j ) ( e^{- \frac{1}{3} \hat x^3} 
e^{\hat x v_i}  - 2 
\int_{0}^{\infty} dy Ai(v_i + y) 
 e^{- \hat x y} )  \nn
\eea 
which is the main result of this paper. It has the form of the ${\cal A}_{2 \to 1}$ transition kernel
plus a projector. 

\subsection{the Brownian-Brownian initial condition}

Consider now the Brownian-Brownian initial condition \eqref{inith},
with $a_L=a_R=1$, i.e. a double sided Brownian initial condition.

\subsubsection{first form of the kernel}

From \eqref{M2} we find:
\bea \label{MBB} 
&& M_s(v_i,v_j) = \int \frac{dk}{2 \pi} dy Ai(y + s+ 4 k^2 + i k \tilde x + v_i + v_j) e^{- 2 i k (v_i-v_j)} 
 \phi_{+\infty}(k,y) \\
&& \phi_{+\infty}(k,y) =  - 2 \theta(y)  - 4 \theta(-y) (e^{(2 \tilde w_L + 2 i k) y} 
+ e^{(2 \tilde w_R - 2 i k) y}) +  \frac{2}{\tilde w_L + \tilde w_R}  \delta(y)
\eea

\subsubsection{second form of the kernel}

The second form of the kernel reads:
\be \label{kernel2BB} 
 K(v_i,v_j) = 
\int_0^{+\infty} dy Ai(v_i + y) Ai(v_j + y)  
- \frac{1 }{\hat w_L + \hat w_R}  Ai(v_i) Ai(v_j)  +  Ai(v_i) {\cal B}_{\hat x + \hat w_L}(v_j)  
+  Ai(v_j) {\cal B}_{-\hat x + \hat w_R}(v_i) 
\ee
where we have defined
\bea
&& {\cal B}_{w}(v) = e^{w^3/3} e^{-v w} - \int_0^\infty dy Ai(v+ y) e^{w y}
= \int_{-\infty}^0 d y Ai(v+ y) e^{w y} \label{defB1} 
\eea 
where the second form is valid for $w>0$, while the 
first one is valid for arbitrary $w$ (see discussion above).

We now show that this result is equivalent to the result of
Ref. \cite{SasamotoStationary} in the large time limit. 
The notations of that paper are $X=\hat x$, $\gamma_t=t^{1/3}$,
$v_\pm=w_{R,L}$ hence $\omega_\pm=v_\pm \gamma_t=\hat w_{R,L}$ 
(with $\alpha=1$ in our units $\nu=1,D=2,\lambda_0=2$). 
The CDF of the height 
was obtained in formula (6.21-22) at at large time (correcting the misprint $X \to -X$ there):
\bea \label{stat2} 
&& F_{\hat w_R,\hat w_L}(\sigma) = (1 + \frac{1}{\hat w_R + \hat w_L} \frac{d}{d\sigma} ) 
{\rm Det}(1 - P_\sigma B P_\sigma) \\
&& B(v_1,v_2) = K_{Ai}(v_1,v_2) + (\hat w_R + \hat w_L) {\cal B}_{\hat w_R - \hat x}(v_1) {\cal B}_{\hat w_L + \hat x}(v_2) 
\eea 
To show that they are the same, let us first express explicitly the derivative
\bea \label{explicitder} 
&& F_{\hat w_R,\hat w_L}(\sigma)
 = {\rm Det}(1 - P_0 B_\sigma P_0) ( 1 -  \frac{1}{\hat w_R + \hat w_L} Tr [(1 - P_0 B_\sigma P_0)^{-1} 
 P_0 \partial_\sigma B_\sigma P_0 ]) 
\eea
we have used that ${\rm Det}(1 - P_\sigma B P_\sigma)={\rm Det}(1 - P_0 B_\sigma P_0)$ 
where $B_\sigma(v_1,v_2)=B(v_1+\sigma,v_2+\sigma)$. To obtain $\partial_\sigma B_\sigma$ we 
calculate the following derivatives
\bea
&& \partial_\sigma K_{Ai}(v_1+\sigma,v_2+\sigma) =  \int_{\lambda>0} \partial_\lambda[ Ai(v_1+\sigma+\lambda) Ai(v_2+\sigma+\lambda) ]
=- Ai(v_1+\sigma) Ai(v_2+\sigma) \\
&& \partial_\sigma {\cal B}_{w}(v+\sigma) = \int_{-\infty}^0 d \lambda \partial_\lambda [Ai(v+\sigma+\lambda)] e^{w \lambda}
= Ai(v+\sigma) -  w {\cal B}_{w}(v+\sigma)
\eea 
where we used integration by parts. 
Hence we obtain
\bea \label{der1} 
&& \frac{1}{\hat w_R + \hat w_L}  \partial_\sigma B_\sigma(v_1,v_2)= - \frac{Ai(v_1+\sigma) Ai(v_2+\sigma)}{\hat w_R + \hat w_L} + Ai(v_1+\sigma) {\cal B}_{\hat w_L+\hat x}(v_2+\sigma) + Ai(v_2+\sigma) {\cal B}_{\hat w_R- \hat x}(v_1+\sigma) \nn
\\
&& - (\hat w_R + \hat w_L) {\cal B}_{\hat w_R- \hat x}(v_1+\sigma) {\cal B}_{\hat w_L+\hat x}(v_2+\sigma)
\eea
Now we note that it can be written as a product
\bea
\frac{1}{\hat w_R + \hat w_L}  \partial_\sigma B_\sigma(v_1,v_2)= - \phi_1(v_1+\sigma) \phi_2(v_2+\sigma)
\eea 
where we have defined 
\bea 
&& 
\phi_1(v_1) =\frac{Ai(v_1)}{\sqrt{\hat w_L + \hat w_R}} - \sqrt{\hat w_L + \hat w_R} {\cal B}_{w_R-\hat x}(v_1) \\
&& \phi_2(v_2) =\frac{Ai(v_2)}{\sqrt{\hat w_L + \hat w_R}} - \sqrt{\hat w_L + \hat w_R} {\cal B}_{w_L+\hat x}(v_2)
\eea
Hence $\frac{1}{\hat w_R + \hat w_L}  \partial_\sigma B_\sigma(v_1,v_2)$ is a projector, which
implies that \eqref{explicitder} can be rewritten as
\bea \label{explicitder2} 
&& F_{\hat w_R,\hat w_L}(\sigma)
 = {\rm Det}(1 - P_0 ( B_\sigma - \phi_1^T \phi_2) P_0) = {\rm Det}(1 - P_0 K_\sigma P_0)
\eea
since one can check that our Kernel (\ref{kernel2BB}) can be written as:
\bea 
&& K(v_1,v_2) = B(v_1,v_2) - \phi_1(v_1) \phi_2(v_2) \\
&&
K_\sigma(v_1,v_2) = B_\sigma(v_1,v_2) - \phi_1(v_1+\sigma) \phi_2(v_2+\sigma)
\eea


As is discussed in Ref. \cite{SasamotoStationary} the expression 
(\ref{stat2}) is equivalent to the one in Theorem 5.1. of Ref. \cite{ImamuraPNG}
derived for the PNG model with external sources.
The relation between (\ref{stat2}) and the result of Baik and Rains in Ref. \cite{png} 
(in terms of the solution of a Painleve II equation) is discussed in Proposition 5.2 of Ref. \cite{SasamotoStationary}. The result also agrees with the one for stationary TASEP \cite{spohnTASEP}
and a rigorous derivation was given in \cite{BCFV}. Note that in 
Ref. \cite{SasamotoStationary} the solution is given for arbitrary time,
which is possible in that case. This provides a test of our more direct (but more empirical) method to
obtain directly the large time limit.

\subsection{Adding a step to the initial condition}

\label{sec:step2} 

\subsubsection{first form of the kernel}

Consider now the step initial condition \eqref{step0}. 
As discussed in previous sections, to obtain the solution for that case,
in the large time limit, we need to calculate the generalized generating function in the large time limit
\bea
\lim_{t \to +\infty} {\rm Prob}\left(t^{-1/3} h(x=2^{-4/3} t^{2/3} \tilde x,t)  < s \right) = g_{+\infty}^\Delta(s) = g_{+\infty}(s_L = s - \tilde \Delta, s_R = s + \tilde \Delta ) 
\eea 
where $\tilde \Delta = \Delta/\lambda$. We will specify to $a_L=a_R=0$, i.e.
step initial condition on top of the wedge. The solutions for the two other
cases, the step plus half-Brownian (or step on top of flat to stationary), and step 
on top of two-sided Brownian are given in Appendices \ref{step-WB} and \ref{step-BB}
respectively.

From \eqref{Mss}, \eqref{ggene2} 
and the result (\ref{phitotal}) in Appendix \ref{app:1} we can write that
\bea
g_{+\infty}(s_L,s_R) = {\rm Det}[ I + P_0 M_{s_L,s_R} P_0 ] 
\eea
Let us give the result for $\tilde \Delta>0$, i.e. $s_R>s_L$
\be
M_{s_L,s_R}(v_i,v_j) = \int \frac{dk}{2 \pi} dy Ai(y + 4 k^2 + i k \tilde x + v_i + v_j) e^{- 2 i k (v_i-v_j)} 
 \phi_{+\infty}(k,y-s_L,y-s_R) 
\ee
with 
\bea
  \phi_{+\infty}(k,y-s_L,y-s_R) &=&  -2  +  2 \theta(s_L-y) 
+ 
2 \theta(y - s_R) e^{- 2 (\tilde w_L +  \tilde w_R) y } e^{2 i k (s_R - s_L)} 
\\
&- &  \frac{\delta(y-s_L)}{\tilde w_L + i k}  
-  \frac{\delta(y-s_R)}{ \tilde w_R -  i k} 
e^{- 2 \tilde w_R  (s_R-s_L)}
e^{2 i k (s_R-s_L)} \nn
\eea 
at this stage we have also kept arbitraty slopes $\tilde w_{L,R}$. 

\subsubsection{second form of the kernel}

We now obtain the second form for the result. The details are given in
Appendix \ref{sec:genstep} and \ref{step-wedge}. We find
\bea \label{formgen2n} 
{\rm Prob}\left(t^{-1/3} (h(x=2 t^{2/3} \hat x,t) + \frac{x^2}{4 t} ) < \sigma \right) = g^{\Delta}_{+\infty}(s) = {\rm Det}[ I -  P_{\sigma-\hat \Delta} K_{\hat \Delta} P_{\sigma-\hat \Delta} ] \quad , \quad \hat \Delta = \Delta/t^{1/3} 
\eea 
with the kernel
\bea \label{step1} 
&& K_{\hat \Delta}(v_i,v_j)= \int_{0}^{+\infty} dy 
Ai(v_i + y  ) 
Ai(v_j+ y  )  - \int _{0}^{+\infty} dy 
 Ai( v_i + y   ) 
Ai(v_j+ y+ 4 \Delta  ) e^{4 \hat x \hat \Delta} \\
&& +  
\int_{0}^{+\infty} dy Ai(v_i + y   ) 
Ai(v_j  - y  ) 
e^{- 2 y  \hat x } +    \int_{0}^{+\infty} dy 
Ai(v_i   -  y    )
Ai(v_j + 4 \hat \Delta +  y  ) 
e^{2 y  \hat x  }
e^{4 \hat x \hat \Delta} \nonumber 
\eea 
The generalization to arbitrary slopes $\hat w_{L,R}>0$ is given in the Appendix, equation \eqref{newMK3}.

Note that \eqref{formgen2n} can also be written, denoting $\sigma'=\sigma - \Delta$, as
\bea \label{formgen3n} 
{\rm Prob}\left(t^{-1/3} (h(x=2 t^{2/3} \hat x,t) - \Delta + \frac{x^2}{4 t} ) < \sigma' \right)  = {\rm Det}[ I -  P_{\sigma'} K_{\hat \Delta} P_{\sigma'} ] 
\eea 
Hence $\sigma'$ measure the fluctuations w.r.t the height level of the step on the left ($x<0$). Thus, 
if $\hat \Delta \to +\infty$ the step size becomes infinite and the height level on the right 
goes to $- \infty$ (relatively to the left). Thus one must find the half-flat kernel, and indeed
one can check that the second and fourth term in \eqref{step1} vanish in that limit, i.e. 
$K_{\hat \Delta \to +\infty}(v_i,v_j) \to K^{\rm half-flat}(v_i,v_j)$, as given in \eqref{K1/2}

In the limit $\hat \Delta \to 0$ the first two terms in \eqref{step1} cancel and one finds
\bea
\label{step11} 
&& K_{\hat \Delta=0^+}(v_i,v_j)= \int_{-\infty}^{+\infty} dy Ai(v_i + y   ) 
Ai(v_j  - y  ) e^{- 2 y  \hat x } = 2^{-1/3} Ai( 2^{-1/3} (v_i+v_j - 2 \hat x^2)) e^{\hat x(v_i-v_j)} 
\eea 
Hence 
\bea
&& K_{\hat \Delta=0^+}(v_i+ \sigma + \hat x^2,v_j+ \sigma + \hat x^2) 
= 2^{-1/3} \tilde K_{GOE}(2^{-1/3} v_i + s/2, 2^{-1/3} v_j + s/2) 
e^{\hat x(v_i-v_j)} \\
&&  \equiv K^{GOE}(v_i+s/2, v_j+s/2) 
\eea 
with $s=2^{2/3} \sigma$ and we recall $K^{GOE}(v_i,v_j) = Ai(v_i+v_j)$. Hence
we recover the result for the flat initial condition \eqref{resflat}.

\section{Conclusion}
\label{sec:conclu} 

In conclusion we have used the replica Bethe ansatz method to study the distribution of the
the scaled interface height at one space time point in the 1D KPZ equation,
with a set of initial conditions which are different on the negative and the positive half line. 
This set contains all standard
crossover classes between respectively flat, droplet and stationary on each side, as well as in presence of 
slopes (i.e. drifts). The method also allows to add a step at the origin for each
of these initial conditions. The slopes and step parameters, as well as the coordinate of the
observation point, are properly scaled with time so that the result is non-trivial in the large time limit
and interpolates between various classes of initial conditions, as they are varied.
This generalizes our previous work on the crossover between flat and droplet. 
In all cases the one point CDF of the height can be expressed as a Fredholm determinant
with various kernels depending on the parameters. All these expressions, although 
obtained starting from the KPZ equation, are conjectured to be universal for all models
in the 1D KPZ class. 

The method contains some heuristics, following previous works,
as it assumes that in the large time limit, a decoupling occurs, so that some terms can be set to unity in the complicated sum over string eigenstates, allowing for an exact calculation. The calculation is performed
by using, and further testing and extending, a combinatorics method introduced by Dotsenko. 
We test the validity of the method in cases where the answer is known, such as flat, droplet and 
their crossover, as well as Brownian and half Brownian. In these cases, it reproduces
the known result, although sometimes naturally leading to new, equivalent, forms for the kernels. 
In all other cases, it produces some conjectures for the kernels. Among them, the flat to stationary crossover kernel is directly obtained. It would be interesting to confirm all the present results by different
methods.

\bigskip

{\it Note added: } while this work was in the last stages of completion, we learned of
the recent work of Quastel and Remenik \cite{QuastelRemenikParabola}, who
obtained a general formula for a very large class
of initial conditions. Although these do not yet allow to average over random initial conditions (such as
Brownian) it would be interesting, in the deterministic case, to compare
their formula (obtained for Airy processes) and the present results
(obtained starting from the KPZ equation). In an even more recent 
work \cite{KPZFixedPoint2} they prove the convergence to such formula
starting from TASEP.

\acknowledgements
I am grateful to A. Borodin, P. Calabrese, I. Corwin, P. Ferrari, J. Quastel and D. Remenik for useful discussions 
and pointing out several important references. 

\appendix

\section{General identity}
\label{app:dots} 

\subsection{Preliminaries} 

Consider the Bethe wave function \eqref{def1}, $\Psi_\mu(X)$, which is a symmetric function of
its arguments, and which, for $x_1\leq ..\leq x_n$ reads
\bea
\Psi_\mu(X) = \sum_{P \in S_n} A_P \,e^{i \sum_\alpha \lambda_{P_\alpha} x_\alpha} \quad , \quad A_P = 
\prod_{1 \leq \alpha < \beta \leq n} a_{\lambda_{P_\beta},\lambda_{P_\alpha}}  \quad , \quad 
a_{\lambda,\lambda'}= 1 + \frac{i }{\lambda - \lambda'} \label{Bd} 
\eea 
where $\lambda_1,\cdots \lambda_n$ are the rapidities. Split the coordinates $x_\alpha$ into two groups, $N_L$ with $\alpha=1, \cdots, n_L$, i.e. 
$x_1 \leq x_2 \leq \cdots \leq x_{n_L}$ and $N_R$ with $\alpha=n_L+1, \cdots, n_L+n_R=n$, 
i.e. $x_{n_L+1} \leq \cdots \leq x_{n}$, a splitting which respects the constraint $x_1\leq \cdots \leq x_n$, 
i.e. such that $x_a \leq x_b$ for all $a \in N_L$ and $b \in N_R$. In \eqref{Bd}, for each permutation $P$ 
a rapidity
$\lambda_{P_\alpha}$ is associated to each coordinate $x_\alpha$, hence for each permutation $P$
a first $n_L$-uplet of rapidities $(\lambda_{P_1}, \cdots, \lambda_{P_{n_L} })$ is associated to the group $N_L$ and
a second $n_R$-uplet, $(\lambda_{P_{n_L+1}}, \cdots, \lambda_{P_{n}})$ is associated to the group $N_R$. \\

In a number of applications one needs to calculate 
\bea \label{needs} 
&& \sum_{P \in S_n}  \prod_{1 \leq \alpha < \beta \leq n}   a_{\lambda_{P_\beta},\lambda_{P_\alpha}} 
F_{n_L}^L[\lambda_{P_1},..\lambda_{P_{n_L}}]  F_{n_R}^R[\lambda_{P_{n_L+1}},..\lambda_{P_{n}}] 
\eea
where here we will consider $F_{n_L}^L$ and $F_{n_R}^R$ to be {\it arbitrary functions} of 
$n_L$, respectively $n_R$, variables, with a priori no symmetry (i.e. functions of the
$n_L$-uplet and $n_R$-uplet, respectively). This is the case for instance
for the calculation of the overlap of $\Psi_\mu(X)$ with any other wave function
which splits into a product over $N_L$ and $N_R$,
see \eqref{nLnR} as an example. Note that there we eventually sum over $n_L,n_R=n-n_L$
but here we will consider the more general question of evaluation of \eqref{needs} for
any fixed $n_L,n_R=n-n_L$. 

Let us consider this question when the rapidities are strings. So 
consider now a Bethe state with $n_s$ strings specified by $k_j,m_j$, $j=1,\cdots,n_s$, i.e. rapidities labeled as:
\bea
\lambda_\alpha \to \lambda_{j,r_j} = k_j - \frac{i}{2} (m_j+1 - 2 r_j) \quad 1 \leq r_j \leq m_j
\eea 
In notations of the text such a state is denoted as $|{\bf k}, {\bf m} \rangle$. 
As is well known, and clear from the definition \eqref{Bd}, the only permutations $P$ which have a non vanishing amplitude $A_P$, are those such that for each string the intra-string order of increasing imaginary part is
maintained. Hence if one is given the set of $2 n_s$ integers $(m^L_j,m^R_j)$, $j=1,\cdots,n_s$:
\bea
&& 0 \leq m_j^L \leq m_j \quad 0 \leq m_j^R \leq m_j \quad m_j^L+m_j^R = m_j 
\eea
which specifies how many particles in each string belongs to each of the two groups,
then one knows (bijectively) the two sets of rapidities which belong of each group.
For instance one knows that the first set of rapidities is:
\bea
 \Lambda_L= \{ \lambda_{j,r_j} , j=1,\cdots,n_s, r_j =1,\cdots,m_j^L \}
\eea
and the second set is the complementary $ \Lambda_R = \{ \lambda_{j,r_j} , j=1,..n_s, r_j =m_j^L+1,..m_j \}$.
To treat these two sets on equal footing, it is convenient to introduce the notation:
\bea
&& \lambda_{j,r_j} = k_j - \frac{i}{2} (m_j+1 - 2 r_j)   \quad 1 \leq r_j \leq m_j^L \\
&& \bar \lambda_{j,r_j} = k_j + \frac{i}{2} (m_j+1 - 2 r_j)   \quad 1 \leq r_j \leq m_j^R
\eea 
Note that the sets are now specified, but that within each set, one still needs to sum over all possible orders, i.e. permutations. That is, 
to each of these two sets, one can associate $n_L!$ possible $n_L$-uplets (respectively $n_R!$ possible $n_R$-uplets)
of rapidities. 

Consider now the quantity defined by \eqref{needs} for a given string state $|{\bf k}, {\bf m} \rangle$.
One can guess that the sum over $P \in S_n$ in \eqref{needs} can now be made in two stages. 
In a first stage one fixes the ${\bf m}^R=\{m^R_j\}_{j=1,\cdots n_s}$, equivalently the 
${\bf m}^L=\{m^L_j\}_{j=1,\cdots n_s}$, and perform the sum over permutations inside
each set, and then, in a second stage sum over the variables ${\bf m}^{R,L}$. 
One takes advantage that one can factor $A_P$ as
\be
\prod_{1 \leq \alpha < \beta \leq n} a_{\lambda_{P_\beta},\lambda_{P_\alpha}}  
= \prod_{1 \leq \alpha < \beta \leq n_L}   a_{\lambda_{P_\beta},\lambda_{P_\alpha}}
\times \prod_{n_L+1 \leq \alpha < \beta \leq n}   a_{\lambda_{P_\beta},\lambda_{P_\alpha}} 
\times \prod_{\alpha =1}^{n_L} \prod_{\beta=n_L+1}^n 
a_{\lambda_{P_\beta},\lambda_{P_\alpha}}  \label{ap1} 
\ee 
and one defines
\bea
&& H_{n_L}^L[\{\lambda_1,..\lambda_{n_L}\}] = \sum_{P \in S_{n_L}} \big( \prod_{1 \leq \alpha < \beta \leq n_L}   a_{\lambda_{P_\beta},\lambda_{P_\alpha}} \big) 
F_{n_L}^L[\lambda_{P_1},..\lambda_{P_{n_L}}]  \\
&& H_{n_R}^R[\{\lambda_{n_L+1},..\lambda_{n}\}] = \sum_{P \in S_{n_R}} \big( \prod_{n_L+1 \leq \alpha < \beta \leq n}   a_{\lambda_{P_\beta},\lambda_{P_\alpha}} \big) 
F_{n_R}^R[\lambda_{P_{n_L+1}},..\lambda_{P_{n}}] 
\eea 
Clearly $H_{n_L}^L$ and $H_{n_R}^R$ are now fully symmetric functions of their arguments. One can
then evaluate $H_{n_L}^L$ on the set $\Lambda_L$ and $H_{n_R}^R$ on the set $\Lambda_R$.
One thus defines:
\bea
&& \tilde H^L[{\bf k}, {\bf m}, {\bf m}^L] = H_{n_L}^L[\lambda_{1,1},..\lambda_{1,m_1^L},..., \lambda_{n_s,1},..
\lambda_{n_s,m_{n_s}^L}] \\
&& \tilde H^R[{\bf k}, {\bf m}, {\bf m}^R] = H_{n_R}^R[\bar \lambda_{1,1},..\bar \lambda_{1,m_1^R},..., \bar \lambda_{n_s,1},..
\bar \lambda_{n_s,m_{n_s}^R}]
\eea 
Note that the functions on the left do not explicitly depend any more on the choice
$n_R,n_L$, they depend on this choice only via $n_L = \sum_j m_j^L$ and 
$n_R = \sum_j m_j^R$. 

Let us now consider the last factor in \eqref{ap1} and evaluate it on the string state $|{\bf k}, {\bf m} \rangle$
\bea
 \prod_{\alpha =1}^{n_L} \prod_{\beta=n_L+1}^n 
a_{\lambda_{P_\beta},\lambda_{P_\alpha}} 
&= & \prod_{1 \leq j < j' \leq n_s} G_{jj'} G_{j'j} \prod_{1 \leq j  \leq n_s} G_{jj} \\
& :=  &  {\cal G}[{\bf k}, {\bf m}^{ L} , {\bf m}^{ R}] 
\eea
where we have defined the function ${\cal G}$ and the factors $G$ can be expressed
using Pochammer symbols $(x)_n = \Gamma(x+n)/\Gamma(x)$ as follows
\bea
&& G_{jj} = \prod_{r=1}^{m_j^L} \prod_{r'=1}^{m_{j}^R} a_{\bar \lambda_{j,r'},\lambda_{j,r}}
= \frac{(m_j^L + m_j^R)!}{m_j^L! m_j^R!} \\
&& G_{jj'} =
 \prod_{r=1}^{m_j^L} \prod_{r'=1}^{m_{j'}^R} a_{\bar \lambda_{j',r'},\lambda_{j,r}} 
= \frac{\left(i \left(k_j-k_{j'}\right)+\frac{1}{2}
   \left(-m_j^L+m_j^R+m_{j'}^L+m_{j'}^R\right)+1\right){}_{m_j^L}}{\left(i \left(k_j-k_{j'}\right)+\frac{1}{2}
   \left(-m_j^L+m_j^R+m_{j'}^L-m_{j'}^R\right)+1\right){}_{m_j^L}} \\
\eea 
where we have replaced everywhere $m_j=m_j^L+m_j^R$,
and $G_{j'j}$ is obtained by simply exchanging all indices $j$ and $j'$. The function ${\cal G}$ can 
thus be written in terms of Gamma functions
\bea \label{Gform1} 
&& {\cal G}[{\bf k}, {\bf m}^{ L} , {\bf m}^{ R}] = \prod_{1 \leq j  \leq n_s} \frac{(m_j^L + m_j^R)!}{m_j^L! m_j^R!} \\
&& \times \prod_{1 \leq j \neq j' \leq n_s}
\frac{\Gamma\left(i \left(k_j-k_{j'}\right)+\frac{1}{2}  \left(m_j^L+m_j^R+m_{j'}^L+m_{j'}^R\right)+1\right)
}{\Gamma\left(i \left(k_j-k_{j'}\right)+\frac{1}{2} \left(-m_j^L+m_j^R+m_{j'}^L+m_{j'}^R\right)+1\right)
}
\frac{\Gamma\left(i \left(k_j-k_{j'}\right)+\frac{1}{2}
   \left(-m_j^L+m_j^R+m_{j'}^L-m_{j'}^R\right)+1\right)
}{\Gamma\left(i \left(k_j-k_{j'}\right)+\frac{1}{2}
   \left(m_j^L+m_j^R+m_{j'}^L-m_{j'}^R\right)+1\right)
} \nonumber 
\eea 
Note that using the identity $\Gamma(1-x)=\pi/(\Gamma(x) \sin(\pi x))$ we can rewrite this
function differently. One can check that for integers $m_j^{L,R}$ the factors containing
the sinus functions all together simplify to unity. Hence the function $G$ can equivalently
be written as
 \bea
{\cal G}= \prod_{1 \leq j \neq j' \leq n_s}  \frac{
\Gamma \left(- i k_j+ i k_{j'}+\frac{m^L{}_j-m^L{}_{j'}-m^R{}_j-m^R{}_{j'}}{2}\right) 
\Gamma \left( i k_j- i k_{j'}+ \frac{ -m^L{}_j-m^L{}_{j'}+m^R{}_j-m^R{}_{j'}}{2}\right)}{
\Gamma \left(- i k_j+ i   k_{j'}- \frac{m^L{}_j+m^L{}_{j'}+m^R{}_j+m^R{}_{j'}}{2}\right) 
\Gamma \left(i k_j- i   k_{j'}+ \frac{ -m^L{}_j+m^L{}_{j'}+m^R{}_j-m^R{}_{j'}}{2}\right)}
\prod_{1 \leq j  \leq n_s} \frac{(m_j^L + m_j^R)!}{m_j^L! m_j^R!} \nonumber 
 \eea 
which shows that the question of its analytic continuation to $m_j^{L,R}$ complex is non-trivial
(non-unique). Indeed if one were to attempt
calculations including this factor using Mellin Barnes formula, one could argue 
from the form \eqref{Gform1} that the standard scaling at large time $k_j \to k_j/\lambda$,
$m_j^{L,R} \to z_j^{L,R}/\lambda$ leads to ${\cal G} \to 1$, however on the second
form such a property does not seem to hold. 

%
%

\subsection{Main identity}

Our main result is the following general identity, for the evaluation of \eqref{needs} for a given string state $|{\bf k}, {\bf m} \rangle$, valid for any fixed
$(n_L,n_R,n_s,m_j,k_j)$ and arbitary functions $F_{n_L}^L$, $F_{n_R}^R$
\bea
&& \sum_{P \in S_n}  \prod_{1 \leq \alpha < \beta \leq n}   a_{\lambda_{P_\beta},\lambda_{P_\alpha}} 
F_{n_L}^L[\lambda_{P_1},..\lambda_{P_{n_L}}]  F_{n_R}^R[\lambda_{P_{n_L+1}},..\lambda_{P_{n}}] \\
&& = \prod_{j=1}^{n_s} \sum_{m_j^L + m_j^R = m_j} \delta_{\sum_{j=1}^{n_s} m_j^L=n_L} 
\delta_{\sum_{j=1}^{n_s} m_j^R=n_R}  \tilde H^L[{\bf k}, {\bf m}, {\bf m}^L]  
\tilde H^R[{\bf k}, {\bf m}, {\bf m}^R] {\cal G}[{\bf k}, {\bf m}^{ L} , {\bf m}^{ R}] 
\eea 
where the functions $\tilde H^L$, $\tilde H^R$ and ${\cal G}$ are given above.
We have not attempted to prove this identity, but we have checked it using mathematica
for a large set of values of the parameters $n_R,n_L,n_s$. 
%
Setting the functions $F_{n_L}^L$, $F_{n_R}^R$ to unity we have also checked the
(quite non-trivial) "normalisation identity":
\bea
 \prod_{j=1}^{n_s} \sum_{m_j^L + m_j^R = m_j} \delta_{\sum_{j=1}^{n_s} m_j^L=n_L} 
\delta_{\sum_{j=1}^{n_s} m_j^R=n_R} {\cal G}[{\bf k}, {\bf m}^{ L} , {\bf m}^{ R}]  = n!/(n_L! n_R!) 
\eea 
which can be seen as an indentity involving Gamma functions. 
We have also checked that the above expression
for ${\cal G}$ (once multiplied by its symmetric) is consistent with the formula given by Dotsenko 
\cite{dotsenkoGOE}. 


\section{Calculation of the auxiliary function $\phi_{+\infty}(k,y_L,y_R)$}
\label{app:1} 

To perform the integrals in (\ref{phi1}) we expand the product, leading to four terms.
We use the elementary integrals
\bea
&& (\frac{-1}{2 \pi i}) \int_{C'} \frac{dz}{z} e^{z y  } = \theta(-y)  \\
&&  (\frac{-1}{2 \pi i}) \int_{C'} \frac{dz}{z} \frac{1}{A +z}  e^{z y  } =
\int_0^{+\infty} dv e^{- A v} \theta(-y+v) =  \frac{1}{A} ( \theta(-y) + \theta(y) e^{- A y})  \quad , \quad {\rm Re}(A)>0 
\eea
This allows to show, assuming everywhere ${\rm Re}(A) >0$:
\bea
&& 
(\frac{-1}{2 \pi i})^2 \int_{C'} \frac{dz_L}{z_L}\int_{C'} 
\frac{dz_R}{z_R}  \frac{e^{ z_L y_L+z_R y_R  }}{A + z_L- a_R z_R} 
=  \int_{v_2>0} e^{-A v_2} \theta(-a_R v_2-y_R) \theta(v_2-y_L) 
\eea
Taking a derivative w.r.t. $y_R$ one obtains:
\bea
&& 
(\frac{-1}{2 \pi i})^2 \int_{C'} \frac{dz_L}{z_L}\int_{C'} 
\frac{dz_R}{z_R}  \frac{z_R e^{ z_L y_L+z_R y_R  }}{A + z_L- a_R z_R} 
= -  \theta(- y_R)  \theta(-y_R-y_L) e^{A y_R} \delta_{a_R,1}
- \delta(y_R) \frac{1}{A} (\theta(-y_L) + \theta(y_L) e^{- A y_L}) \delta_{a_R,0} \nn
\eea
which allows to evaluate the two cross-terms in (\ref{phi1}).
We also need:
\bea
&& (\frac{-1}{2 \pi i})^2 \int_{C'} \frac{dz_L}{z_L}\int_{C'} 
\frac{dz_R}{z_R}  \frac{e^{ z_L y_L+z_R y_R  }}{(A_L+ z_R- a_L z_L)(A_R + z_L-a_R z_R)} \\
&& 
= \int_{v_1>0,v_2>0} e^{-A_L v_1- A_R v_2} \theta(v_1-y_R-a_R v_2) \theta(v_2-y_L-a_L v_1) \nn
\eea
and taking two derivatives we obtain:
\bea
&&
(\frac{-1}{2 \pi i})^2 \int_{C'} \frac{dz_L}{z_L}\int_{C'} 
\frac{dz_R}{z_R}  \frac{z_L z_R e^{ z_L y_L+z_R y_R  }}{(A_L+ z_R- a_L z_L)(A_R+ z_L-a_R z_R)} 
\\
&& = \theta(y_L+ a_L y_R) \theta(y_R+ a_R y_L) 
e^{- A_L (y_R+ a_R y_L) - A_R (y_L+ a_L y_R)}  (1 - a_R a_L) \nn \\
&& + a_R a_L 
\delta(y_L + y_R) \frac{1}{A_L+A_R} (\theta(y_L) e^{- A_R y_L} + \theta(-y_L) e^{A_L y_L} )  \nn
\eea 

Putting all together, denoting $A_L=2 \tilde w_L + 2 i k$ and $A_R=2 \tilde w_R - 2 i k$ 
and slightly simplifying using that $(a_L,a_R) \in \{0,1\}^2$ we obtain
from (\ref{phi1})
\bea \label{phitotal} 
&& \frac{1}{2} \phi_{+\infty}(k,y_L,y_R) = -1 +  \theta(-y_L) \theta(-y_R)  
\\
&& - 
   2 a_L  \theta(- y_L)  \theta(-y_R-y_L) e^{A_L y_L} 
- (1-a_L) \delta(y_L) \frac{1}{A_L} (\theta(-y_R) + \theta(y_R) e^{- A_L y_R})  \nn \\
&& - 
    2 a_R \theta(- y_R)  \theta(-y_R-y_L) e^{A_R y_R}
- (1-a_R) \delta(y_R) \frac{1}{A_R} (\theta(-y_L) + \theta(y_L) e^{- A_R y_L})  \nn \\
&& +  (1+ a_L+ a_R - 3 a_R a_L)
\theta(y_L+ a_L y_R) \theta(y_R+ a_R y_L) 
e^{- A_L (y_R+ a_R y_L) - A_R (y_L+ a_L y_R)}  \nn \\
&& + 4 a_R a_L 
\delta(y_L + y_R) \frac{1}{A_L+A_R} (\theta(y_L) e^{- A_R y_L} + \theta(-y_L) e^{A_L y_L} )  \nn
\eea 
If we set $y_L=y_R=y$ we obtain the formula \eqref{phi2} in the text.

\section{Airy function identities and second form of the Kernel}
\label{app:Airy} 

We use the Airy function identities (see e.g. Section 9 in Ref. \cite{PLDCrossoverDropFlat} and
references therein)
\bea
&& 2 \int \frac{dk}{2 \pi} Ai(4 k^2 + a + b + i k \tilde x) e^{2 i k (b-a)} = 2^{-1/3} Ai(2^{1/3} (a + \frac{\tilde x^2}{32}) )
Ai(2^{1/3} (b + \frac{\tilde x^2}{32})) e^{\frac{\tilde x}{4}(b-a)}  \\
&& 2 \int \frac{dk}{2 \pi} Ai(4 k^2 + a + b + i k \tilde x)  \frac{e^{2 i k (b-a)}}{\tilde w + i k} \\
&& = 
\int_{0}^{+\infty} dr 2^{-1/3} Ai(2^{1/3} (a + \frac{r}{4} +  \frac{\tilde x^2}{32}) )
Ai(2^{1/3} (b - \frac{r}{4} + \frac{\tilde x^2}{32})) e^{\frac{\tilde x}{4}(b-a) - r ( \frac{\tilde x}{8} + \tilde w) } \\
&& 2 \int \frac{dk}{2 \pi} Ai(4 k^2 + a + b + i k \tilde x)  \frac{e^{2 i k (b-a)}}{\tilde w - i k} \\
&& = 
\int_{0}^{+\infty} dr 2^{-1/3} Ai(2^{1/3} (b + \frac{r}{4} +  \frac{\tilde x^2}{32}) )
Ai(2^{1/3} (a - \frac{r}{4} + \frac{\tilde x^2}{32})) e^{\frac{\tilde x}{4}(b-a) - r ( - \frac{\tilde x}{8} + \tilde w) }
\eea 
where we assumed $\tilde w>0$. 

Consider now the expression (\ref{M2}) for the kernel $M_s(v_i,v_j)$ 
and enumerate the terms upon expanding the products in (\ref{phi2}).
In the same order as they appear there, we use the above identities as follows. In the first four terms
we use the first identity. In the
first term we use $a=v_i + (y+s)/2$ and $b=v_j+ (y+s)/2$, in the second term we use
$a=v_i +s/2 + y (1-a_L+a_R)/2$ and $b=v_j+ s/2 + y (1+a_L-a_R)/2$, in the
third term we use $a=v_i +s/2$ and $b=v_j+ s/2 + y$, in the fourth term
we use $a=v_i +s/2+y$ and $b=v_j+ s/2$. In the last three terms we 
use the same $a,b$ as in the first term, and use respectively the second, third and first
identities. This gives $M_s(v_i,v_j) = e^{\frac{\tilde x}{4}(v_j-v_i)} \tilde M_s(v_i,v_j)$ with:
\bea
&& \tilde M_s(v_i,v_j) = - \int dy \bigg(  \theta(y) 
2^{-1/3} Ai(2^{1/3} ( v_i + \frac{y+s}{2} + \frac{\tilde x^2}{32}) ) Ai(2^{1/3} (v_j+ \frac{y+s}{2} + \frac{\tilde x^2}{32}))   \\
&& -  2^{-1/3} Ai(2^{1/3} (v_i +\frac{s}{2} + \frac{(1-a_L+a_R) y}{2} + \frac{\tilde x^2}{32}) ) Ai(2^{1/3} (v_j +\frac{s}{2} 
+  \frac{(1+a_L-a_R) y}{2} + \frac{\tilde x^2}{32})) \\
&& \times \theta(y) (1+a_L+a_R-3 a_L a_R) e^{\frac{\tilde x}{4} y(a_L-a_R)} 
 e^{- 2 (\tilde w_L  (1+ a_R) + \tilde w_R (1+ a_L)) y} \\
&& + 2 \theta(-y) a_L 2^{-1/3}  Ai(2^{1/3} (v_i + \frac{s}{2} + \frac{\tilde x^2}{32}) ) Ai(2^{1/3} (v_j+ \frac{s}{2} + y + \frac{\tilde x^2}{32})) e^{\frac{\tilde x}{4} y } e^{2 \tilde w_L y}  \\
&& + 2 \theta(-y) a_R 2^{-1/3} Ai(2^{1/3} (v_i + \frac{s}{2} + y  + \frac{\tilde x^2}{32}) ) 
Ai(2^{1/3} (v_j+ \frac{s}{2}  + \frac{\tilde x^2}{32})) e^{- \frac{\tilde x}{4} y} e^{2 \tilde w_R y}  \\
&& + \frac{1}{2} (1-a_L) \delta(y)
\int_{0}^{+\infty} dr 2^{-1/3} Ai(2^{1/3} (v_i + \frac{s}{2} + \frac{r}{4} +  \frac{\tilde x^2}{32}) )
Ai(2^{1/3} (v_j+ \frac{s}{2} - \frac{r}{4} + \frac{\tilde x^2}{32})) 
e^{ - r ( \frac{\tilde x}{8} + \tilde w_L) } \\
&& + \frac{1}{2} (1-a_R) \delta(y)
\int_{0}^{+\infty} dr 2^{-1/3} Ai(2^{1/3} (v_i + \frac{s}{2} - \frac{r}{4} +  \frac{\tilde x^2}{32}) )
Ai(2^{1/3} (v_j+ \frac{s}{2} + \frac{r}{4} + \frac{\tilde x^2}{32})) 
e^{ - r ( - \frac{\tilde x}{8} + \tilde w_R) } \\
&& - \delta(y) \frac{a_R a_L }{\tilde w_L + \tilde w_R} 
2^{-1/3} Ai(2^{1/3} ( v_i + \frac{s}{2} + \frac{\tilde x^2}{32}) ) Ai(2^{1/3} (v_j+ \frac{s}{2} + \frac{\tilde x^2}{32}))   \bigg)
\nn
\eea
In the final Fredholm determinant the common factor $e^{\frac{\tilde x}{4}(v_j-v_i)}$ 
can be discarded, since ${\rm Det}[I + P_0 M_s P_0 ] = {\rm Det}[I + P_0 \tilde M_s P_0]$.

We now rescale $y \to 2^{2/3} y$ in the first and second term,
$y \to 2^{-1/3} y$ in the third and fourth term, and $r \to 2^{5/3} y$ in the 
last two terms, and we
use the similarity transformation
$M_s(v_1,v_2) = - 2^{1/3} K_\sigma(2^{1/3} v_1, 2^{1/3} v_2)$ 
and we obtain the result (\ref{final1},\ref{final2}) displayed in the text.

\section{Generalized kernels and step initial conditions}
\label{sec:genstep} 

\subsection{General case: first form of kernel} 

Here we obtain the kernels associated to the generalized generating function \eqref{genfunctLR}.
We start with the first form \eqref{Mss}-\eqref{ggene2} 
\bea
g_{+\infty}(s_L,s_R) = {\rm Det}[ I + P_0 M_{s_L,s_R} P_0 ] 
\eea
To express $\phi_{+\infty}(k,y-s_L,y-s_R)$ let us insert $y_L=y-s_L$ and $y_R=y-s_R$ in \eqref{phitotal}. 
We obtain
\bea \label{M10}
 && M_{s_L,s_R}(v_i,v_j) = \int \frac{dk}{2 \pi} dy Ai(y + 4 k^2 + i k \tilde x + v_i + v_j) e^{- 2 i k (v_i-v_j)} 
 \phi_{+\infty}(k,y-s_L,y-s_R) \\
&& \frac{1}{2} \phi_{+\infty}(k,y-s_L,y-s_R) = -1 +  \theta(s_L-y) \theta(s_R-y)  \label{M11}
\\
&& - 
   2 a_L  \theta(s_L- y)  \theta(s_L+s_R-2 y) e^{A_L (y-s_L)} 
- (1-a_L) \delta(y-s_L) \frac{1}{A_L} (\theta(s_R-s_L) + \theta(s_L-s_R) e^{- A_L (s_L-s_R)})  \nn \\
&& - 
    2 a_R \theta(s_R- y)  \theta(s_L+s_R-2 y) e^{A_R (y-s_R)}
- (1-a_R) \delta(y-s_R) \frac{1}{A_R} (\theta(s_L-s_R) + \theta(s_R-s_L) e^{- A_R (s_R-s_L)}) \nn \\
&& +  (1+ a_L+ a_R - 3 a_R a_L)
\theta((1+a_L)y - s_L- a_L s_R) \theta((1+a_R) y - s_R- a_R s_L) \nn \\
&& \times 
e^{- A_L ((1+a_R) y - s_R- a_R s_L) - A_R ((1+a_L) y - s_L- a_L s_R)} \nn  \\
&& + 2 a_R a_L 
\delta(y - \frac{s_L + s_R}{2}) \frac{1}{A_L+A_R} (\theta(s_R-s_L) e^{- \frac{1}{2} A_R (s_R-s_L)} 
+ \theta(s_L-s_R) e^{\frac{1}{2} A_L (s_R-s_L)} )  \nn
\eea 
with $A_L=2 \tilde w_L + 2 i k$ and $A_R=2 \tilde w_R - 2 i k$. 

\subsection{General case: second form of kernel} 

We now rewrite the kernel using the
Airy function identities given in the previous section. 
This gives $M_{s_L,s_R}(v_i,v_j) = e^{\frac{\tilde x}{4}(v_j-v_i)} \tilde M_s(v_i,v_j)$ with:
\bea \label{newM} 
&& \tilde M_{s_L,s_R}(v_i,v_j) = - \int dy \bigg(  (1  - \theta(s_L-y) \theta(s_R-y) )
2^{-1/3} Ai(2^{1/3} ( v_i + \frac{y}{2} + \frac{\tilde x^2}{32}) ) 
Ai(2^{1/3} (v_j+ \frac{y}{2} + \frac{\tilde x^2}{32}))   \\
&& - (1+ a_L+ a_R - 3 a_R a_L)
\theta((1+a_L)y - s_L- a_L s_R) \theta((1+a_R) y - s_R- a_R s_L) \nn \\
&&
\times e^{(2 \tilde w_L +\frac{\tilde x}{4}) (s_R+a_R s_L) + (2 \tilde w_R -\frac{\tilde x}{4}) ( s_L + a_L s_R)}
e^{- 2 (\tilde w_L(1+a_R) +  \tilde w_R(1+a_L) ) y + \frac{\tilde x}{4} y (a_L-a_R)}  \nn \\
&&
\times 2^{-1/3} Ai(2^{1/3} ( v_i + \frac{1}{2} (y(1-a_L+a_R)+s_L(1-a_R)-s_R(1-a_L)) + \frac{\tilde x^2}{32}) ) \nn \\
&& \times 
Ai(2^{1/3} (v_j+ \frac{1}{2} (y(1+a_L-a_R)-s_L(1-a_R)+s_R(1-a_L))  + \frac{\tilde x^2}{32}))  \nonumber \\
&&+ 2 a_L  \theta(s_L- y)  \theta(s_L+s_R-2 y) e^{(2 \tilde w_L+\frac{\tilde x}{4})(y-s_L)} 
2^{-1/3} Ai(2^{1/3} ( v_i + \frac{s_L}{2} + \frac{\tilde x^2}{32}) ) 
Ai(2^{1/3} (v_j+ y-\frac{s_L}{2} + \frac{\tilde x^2}{32})) \nn \\
&&+ 2 a_R  \theta(s_R- y)  \theta(s_L+s_R-2 y) e^{(2 \tilde w_R-\frac{\tilde x}{4})(y-s_R)} 
2^{-1/3} Ai(2^{1/3} ( v_i + y - \frac{s_R}{2} + \frac{\tilde x^2}{32}) ) 
Ai(2^{1/3} (v_j+\frac{s_R}{2} + \frac{\tilde x^2}{32}))\nn  \\
&& + \frac{1-a_L}{2}  \delta(y-s_L)  \theta(s_R-s_L)
\int_{0}^{+\infty} dr 2^{-1/3} Ai(2^{1/3} (v_i + \frac{y}{2} + \frac{r}{4} +  \frac{\tilde x^2}{32}) )
Ai(2^{1/3} (v_j + \frac{y}{2} - \frac{r}{4} + \frac{\tilde x^2}{32})) 
e^{- r ( \frac{\tilde x}{8} + \tilde w_L) } \nonumber \\
&& 
+ \frac{1-a_L}{2}  \delta(y-s_L) \theta(s_L-s_R) 
e^{- (2 \tilde w_L+\frac{\tilde x}{4})  (s_L-s_R)} \nn
\\
&& \times \int_{0}^{+\infty} dr 2^{-1/3} Ai(2^{1/3} (v_i + \frac{y + s_L-s_R}{2} + \frac{r}{4} +  \frac{\tilde x^2}{32}) )
Ai(2^{1/3} (v_j+ \frac{y+s_R-s_L}{2} - \frac{r}{4} + \frac{\tilde x^2}{32})) 
e^{- r ( \frac{\tilde x}{8} + \tilde w_L) }  \nonumber 
\\
&&  + \frac{1-a_R}{2}  \delta(y-s_R)  \theta(s_L-s_R)
\int_{0}^{+\infty} dr 2^{-1/3} Ai(2^{1/3} (v_i + \frac{y}{2} - \frac{r}{4} +  \frac{\tilde x^2}{32}) )
Ai(2^{1/3} (v_j + \frac{y}{2} + \frac{r}{4} + \frac{\tilde x^2}{32})) 
e^{- r ( - \frac{\tilde x}{8} + \tilde w_R) } \nonumber \\
&& + \frac{1-a_R}{2}  \delta(y-s_R) \theta(s_R-s_L) 
e^{- (2 \tilde w_R - \frac{\tilde x}{4})(s_R-s_L)}
\\
&& \times \int_{0}^{+\infty} dr 2^{-1/3} Ai(2^{1/3} (v_i + \frac{y + s_L-s_R}{2} - \frac{r}{4} +  \frac{\tilde x^2}{32}) )
Ai(2^{1/3} (v_j+ \frac{y+s_R-s_L}{2} + \frac{r}{4} + \frac{\tilde x^2}{32})) 
e^{- r ( - \frac{\tilde x}{8} + \tilde w_R) }  \nonumber \\
&&  - \frac{a_R a_L}{\tilde w_L + \tilde w_R}
\delta(y - \frac{s_L + s_R}{2}) [ \theta(s_R-s_L) 
e^{- \frac{1}{2} (2 \tilde w_R - \frac{\tilde x}{4})(s_R-s_L)} + \theta(s_L-s_R)  e^{\frac{1}{2} (2 \tilde w_L + \frac{\tilde x}{4})(s_R-s_L)} ] \nn \\
&& \times  2^{-1/3} Ai(2^{1/3} (v_i + \frac{y}{2} + \frac{s_L-s_R}{4} +  \frac{\tilde x^2}{32}) )
Ai(2^{1/3} (v_j+ \frac{y}{2}+\frac{s_R-s_L}{4} + \frac{\tilde x^2}{32})) 
 \nonumber 
\bigg)
\eea
a sum of ten terms. 
The identities are used with $a=v_i + \frac{y-R}{2}$ and $b=v_j+\frac{y+R}{2}$
where $R$ is as follows. In term $1$ of \eqref{newM} (which comes from 
term $1,2$ of \eqref{M11}) we use $R=0$ and identity $1$, then we list
similarly: term $2$ (term 7) $R=y(a_L-a_R)+s_R(1-a_L) - s_L(1-a_R)$ identity $1$;
term $3$ (term 3) $R=y-s_L$ identity $1$;
term $4$ (term 5) $R=s_R-y$ identity $1$;
term $5$ (first piece of term 4) $R=0$ identity $2$;
term $6$ (second piece of term 4) $R=s_R-s_L$ identity $2$;
term $7$ (first piece of term 6) $R=0$ identity $3$;
term $8$ (second piece of term 6) $R=s_R-s_L$ identity $3$;
term $9$ (first piece of term 8) $R=\frac{1}{2} (s_R-s_L)$ identity $1$;
term $10$ (second piece of term 8) $R=\frac{1}{2} (s_R-s_L)$ identity $1$;

We now want rewrite the generating function using the second kernel
\bea
g_{+\infty}(s_L,s_R) = {\rm Det}[ I - P_0 K_{\sigma_L,\sigma_R} P_0 ] 
\eea
and we define
\bea
\sigma_L - \hat x^2 = 2^{-2/3} s_L  \quad , \quad \sigma_R - \hat x^2= 2^{-2/3} s_R 
\eea 
where we use that $2^{1/3}  \frac{\tilde x^2}{32} = \hat x^2$. We obtain
\bea \label{newMK} 
&& K_{\sigma_L,\sigma_R}(v_i,v_j) = \int dy \, (1  - \theta(\sigma_L-y) \theta(\sigma_R-y) ) 
Ai(v_i + y) 
Ai(v_j+ y) \\
&& - (1+ a_L+ a_R - 3 a_R a_L) \int dy \,
\theta((1+a_L)y - \sigma_L- a_L \sigma_R) \theta((1+a_R) y - \sigma_R- a_R \sigma_L) \nn \\
&&
\times e^{2 (\hat w_L + \hat x) (\sigma_R+a_R \sigma_L) + 2 (\hat w_R -\hat x) ( \sigma_L + a_L \sigma_R)}
e^{- 2 (\hat w_L(1+a_R) +  \hat w_R(1+a_L) ) y + 2 \hat x y (a_L-a_R)}  \nn \\
&&
\times Ai( v_i + y(1-a_L+a_R)+\sigma_L(1-a_R)-\sigma_R(1-a_L))  
Ai(v_j+ y(1+a_L-a_R)-\sigma_L(1-a_R)+\sigma_R(1-a_L))  \nonumber \\
&&+ 2 a_L \int dy \,  \theta(\sigma_L- y)  \theta(\sigma_L+\sigma_R-2 y) e^{2 (\hat w_L+\hat x)(y-\sigma_L)} 
Ai(v_i + \sigma_L) 
Ai(v_j+ 2 y- \sigma_L) \nn \\
&&+ 2 a_R  \int dy \,  \theta(\sigma_R- y)  \theta(\sigma_L+\sigma_R-2 y) e^{2 (\hat w_R-\hat x)(y-\sigma_R)} 
Ai(v_i + 2 y - \sigma_R) 
Ai(v_j+ \sigma_R)\nn  \\
&& + (1-a_L)   \theta(\sigma_R-\sigma_L)
\int_{0}^{+\infty} dy Ai(v_i + \sigma_L + y)
Ai(v_j + \sigma_L - y) 
e^{- 2 y (\hat w_L + \hat x) } \nonumber \\
&& 
+ (1-a_L)   \theta(\sigma_L-\sigma_R) 
e^{- 2 (\hat w_L+\hat x)  (\sigma_L-\sigma_R)}  \int_{0}^{+\infty} dy Ai(v_i + 2 \sigma_L-\sigma_R  + y)
Ai(v_j+ \sigma_R - y) 
e^{- 2 y ( \hat w_L + \hat x) }  \nonumber 
\\
&&  + (1-a_R)   \theta(\sigma_L-\sigma_R)
\int_{0}^{+\infty} dy Ai(v_i + \sigma_R - y)
Ai(v_j + \sigma_R + y)  
e^{- 2 y (  \hat w_R - \hat x) } \nonumber \\
&& + (1-a_R)  \theta(\sigma_R-\sigma_L) 
e^{- 2 (\hat w_R - \hat x)(\sigma_R-\sigma_L)} \int_{0}^{+\infty} dy Ai(v_i + \sigma_L - y) 
Ai(v_j+ 2 \sigma_R- \sigma_L + y) 
e^{- 2 y (\hat w_R - \hat x) }  \nonumber \\
&&  - \frac{a_R a_L}{\hat w_L + \hat w_R}
 [ \theta(\sigma_R-\sigma_L) 
e^{- (\hat w_R - \hat x)(\sigma_R-\sigma_L)} + \theta(\sigma_L-\sigma_R)  
e^{(\hat w_L + \hat x)(\sigma_R-\sigma_L)} ] Ai(v_i +  \sigma_L)  
Ai(v_j+\sigma_R) 
 \nonumber 
\eea
To obtain this it it more convenient to first define $\sigma_{L,R} = 2^{-2/3} s_L$.
Then, in all terms we have performed a similarity transformation $2^{1/3} v_i, 2^{1/3} v_j \to v_i , v_j$
which multiplies the kernel by $2^{-1/3}$. In terms $1-4$ we have rescaled $y \to 2^{2/3} y$,
in terms $5-8$ we have integrated over the delta functions, then changed variable $r=2^{5/3} y$.
The last step was to make the substitution in the resulting formula, $\sigma_{L,R} \to 
\sigma_{L,R} - \hat x^2$ and, simultaneously change $y \to y - \hat x^2$ but only in terms
$1-4$. 

\subsection{Step on top of the wedge initial condition} 
\label{step-wedge} 

We now want to apply this formula to the step initial conditions. From the text we have
\bea
\lim_{t \to +\infty} {\rm Prob}\left(t^{-1/3} h(x=2^{-4/3} t^{2/3} \tilde x,t)  < s \right) = g_{+\infty}^\Delta(s) = g_{+\infty}(s_L = s - \tilde \Delta, s_R = s + \tilde \Delta ) 
\eea 
where $\tilde \Delta = \Delta/\lambda$. This can be rewritten using the above results
as
\bea
\lim_{t \to +\infty} {\rm Prob}\left(t^{-1/3} ( h(x=2 t^{2/3} \hat x,t) + \frac{x^2}{4 t}) < \sigma \right) = 
g_{+\infty}^\Delta(s) =
 {\rm Det}[ I - P_0 K_{\sigma_L=\sigma- \hat \Delta,\sigma_R=\sigma+ \hat \Delta} P_0 ] 
\eea  
where we recall $\hat \Delta = \Delta/t^{1/3}$.

Let us specify to the case $a_L=a_R=0$, which represents the wedge plus a step. 
With no loss of generality, let us restrict to the case $\sigma_R > \sigma_L$, 
i.e. $\Delta>0$. The kernel then can be written
\bea \label{newMK3} 
&& K_{\sigma_L,\sigma_R}(v_i,v_j) = K_{\hat \Delta = (\sigma_R-\sigma_L)/2}(v_i+\sigma_L,v_j+\sigma_L) \\
&& K_{\hat \Delta}(v_i,v_j)= \int_0^{+\infty} dy \, 
Ai(v_i + y) [ 
Ai(v_j+ y) - e^{4 \hat \Delta (\hat x - \hat w_R)} 
e^{- 2 y (\hat w_L +  \hat w_R ) }  
Ai(v_j+ y + 4 \hat \Delta) ]  \nonumber \\
&& +   
\int_{0}^{+\infty} dy Ai(v_i  + y)
Ai(v_j - y) 
e^{- 2 y (\hat w_L + \hat x) } +  
e^{4 \hat \Delta (\hat x - \hat w_R)} \int_{0}^{+\infty} dy Ai(v_i - y) 
Ai(v_j+ 4 \hat \Delta + y) 
e^{- 2 y (\hat w_R - \hat x) }   \nonumber 
\eea

We can now consider the limit
$\hat w_{L,R} \to 0^+$, which leads to the well defined (trace class)
kernel given in \eqref{step1}. 

\subsection{Step on top of the Brownian-Brownian initial condition}  
\label{step-BB} 

Let us specify to the case $a_L=a_R=1$, which represents the two-sided Brownian (plus drifts)
initial condition plus a step. 
With no loss of generality, let us restrict to the case $\sigma_R > \sigma_L$, 
i.e. $\Delta>0$. The kernel then can be written
\bea \label{newMK5} 
&& K_{\sigma_L,\sigma_R}(v_i,v_j) = K_{\hat \Delta = (\sigma_R-\sigma_L)/2}(v_i+\sigma_L,v_j+\sigma_L) \\
&& K_{\hat \Delta}(v_i,v_j)= \int_0^{+\infty} dy   
Ai(v_i + y) 
Ai(v_j+ y) +  \int_{-\infty}^0 dy \,    e^{ (\hat w_L+\hat x)y } 
Ai(v_i) 
Ai(v_j+ y) \nn \\
&&+    \int_{-\infty}^0 dy  \,  e^{ (\hat w_R-\hat x)(y- 2 \hat \Delta)} 
Ai(v_i +  y) 
Ai(v_j+ 2 \hat \Delta)  - \frac{1}{\hat w_L + \hat w_R}
e^{- 2 \hat \Delta (\hat w_R - \hat x)} Ai(v_i)  
Ai(v_j+ 2 \hat \Delta) 
 \nonumber 
\bigg)
\eea
It can be rewritten is a more generally valid form
\be \label{SBB} 
K_{\hat \Delta}(v_i,v_j)= K_{\rm Ai}(v_i,v_j) - \frac{e^{- 2 \hat \Delta (\hat w_R - \hat x)} }{\hat w_L + \hat w_R}
Ai(v_i)  Ai(v_j+ 2 \hat \Delta) + Ai(v_i) {\cal B}_{\hat w_L + \hat x}(v_j) 
+ e^{- 2 \hat \Delta (\hat w_R - \hat x)} Ai(v_j+ 2 \hat \Delta) {\cal B}_{\hat w_R - \hat x}(v_i) 
\ee
using the functions ${\cal B}_{w}(v)$ defined in (\ref{defB1}). On this form
it is apparent that as $\hat \Delta \to 0$ the kernel converges to the one for
the Brownian-Brownian case \eqref{kernel2BB}. In the opposition limit 
$\hat \Delta \to +\infty$ we see that it converges as it should to the half-Brownian
limit \eqref{hb1} (upon exchange of left and right, and kernel transposition).

\subsection{Step on top of the wedge-Brownian initial condition}  
\label{step-WB} 

Let us specify to the case $a_L=0$, $a_R=1$, which represents an initial condition
which is flat on the left (with a drift), Brownian on the right (with a drift) and, on
top of it, a step. We obtain, from \eqref{newMK}
\bea \label{newMK15} 
&& K_{\sigma_L,\sigma_R}(v_i,v_j) = \int dy \, (1  - \theta(\sigma_L-y) \theta(\sigma_R-y) ) 
Ai(v_i + y) 
Ai(v_j+ y) \\
&& - 2  \int dy \,
\theta(y - \sigma_L) \theta(2 y - \sigma_R-  \sigma_L) e^{2 (\hat w_L + \hat x) (\sigma_R+ \sigma_L) + 2 (\hat w_R -\hat x)  \sigma_L}
e^{- 2 (2 \hat w_L +  \hat w_R ) y - 2 \hat x y}  \nn \\
&&
\times Ai( v_i + 2 y -\sigma_R)  
Ai(v_j +\sigma_R)  \nonumber \\
&&+ 2   \int dy \,  \theta(\sigma_R- y)  \theta(\sigma_L+\sigma_R-2 y) e^{2 (\hat w_R-\hat x)(y-\sigma_R)} 
Ai(v_i + 2 y - \sigma_R) 
Ai(v_j+ \sigma_R)\nn  \\
&& +    \theta(\sigma_R-\sigma_L)
\int_{0}^{+\infty} dy Ai(v_i + \sigma_L + y)
Ai(v_j + \sigma_L - y) 
e^{- 2 y (\hat w_L + \hat x) } \nonumber \\
&& 
+    \theta(\sigma_L-\sigma_R) 
e^{- 2 (\hat w_L+\hat x)  (\sigma_L-\sigma_R)}  \int_{0}^{+\infty} dy Ai(v_i + 2 \sigma_L-\sigma_R  + y)
Ai(v_j+ \sigma_R - y) 
e^{- 2 y ( \hat w_L + \hat x) }  \nonumber 
\eea
Now we must distinguish the two cases $\Delta>0$ and $\Delta<0$.

Let us start with $\Delta>0$ (downward step), i.e. the case $\sigma_R > \sigma_L$.
The kernel then can be written
\bea \label{newMK16} 
&& K_{\sigma_L,\sigma_R}(v_i,v_j)  = K_{\hat \Delta = (\sigma_R-\sigma_L)/2}(v_i+\sigma_L,v_j+\sigma_L) \\
&& K_{\hat \Delta}(v_i,v_j)= \int_{0}^{+\infty} dy \, 
Ai(v_i + y) 
Ai(v_j + y) -   \int_{0}^{+\infty} dy \,
 e^{2 (\hat x-\hat w_R) \hat \Delta}
e^{-  (2 \hat w_L +  \hat w_R ) y -  \hat x y}  Ai( v_i +  y)  
Ai(v_j  + 2 \hat \Delta)  \nonumber \\
&&+    \int_{-\infty}^0  e^{ (\hat w_R-\hat x)(y- 2 \hat \Delta)} 
Ai(v_i +  y ) 
Ai(v_j+ 2 \hat \Delta) +   
\int_{0}^{+\infty} dy Ai(v_i  + y)
Ai(v_j  - y) 
e^{- 2 y (\hat w_L + \hat x) } \nonumber 
\eea
For $\hat \Delta=+\infty$ it goes as expected to the half flat kernel given in \eqref{K1/2}.
For $\hat \Delta=0^+$ is goes to the wedge-Brownian kernel given in \eqref{wb2}. 

Let us consider now $\Delta<0$ (upward step), i.e. the case $\sigma_R < \sigma_L$.
The kernel then can be written
\bea \label{newMK17} 
&& K_{\sigma_L,\sigma_R}(v_i,v_j)  
= \tilde K_{\hat \Delta = (\sigma_R-\sigma_L)/2}(v_i+\sigma_R,v_j+\sigma_R) \\
&& \tilde K_{\hat \Delta}(v_i,v_j) = \int_0^{+\infty} dy \, 
Ai(v_i  + y) 
Ai(v_j + y)  -   \int_0^{+\infty} dy \,
 e^{4 (\hat w_L + \hat x) \hat \Delta}
e^{-  (2 \hat w_L +  \hat w_R ) y -  \hat x y}  Ai( v_i +  y  - 4 \hat \Delta)  
Ai(v_j)  \nonumber \\
&&+    \int_{-\infty}^0 dy \,  e^{ (\hat w_R-\hat x) y} 
Ai(v_i + y ) 
Ai(v_j) +    
e^{4 \hat \Delta (\hat w_L+\hat x)}  \int_{0}^{+\infty} dy Ai(v_i + \sigma_R - 4 \hat \Delta   + y)
Ai(v_j+ \sigma_R - y) 
e^{- 2 y ( \hat w_L + \hat x) }  \nonumber 
\eea
In the limit $\hat \Delta \to -\infty$ we see that it converges as it should to the half-Brownian
limit \eqref{hb1}, and for $\hat \Delta=0^-$ is goes to the wedge-Brownian kernel given in \eqref{wb2},
being of course continuous at $\hat \Delta=0$.

\end{document}